\documentclass[10pt, journal]{IEEEtran}
\usepackage{comment}
\usepackage[table]{xcolor}
\usepackage{cite}
\ifCLASSINFOpdf
   \usepackage[pdftex]{graphicx}
   \DeclareGraphicsExtensions{.pdf,.jpeg,.png}
\fi

\usepackage{amsmath,amssymb}
\usepackage{algorithmic}
\usepackage{array}
\usepackage{stfloats} 
\usepackage{pifont}
\usepackage{booktabs}

\ifCLASSOPTIONcompsoc
  \usepackage[caption=false,font=normalsize,labelfont=sf,textfont=sf]{subfig}
\else
  \usepackage[caption=false,font=footnotesize]{subfig}
\fi

\usepackage{stfloats}
\usepackage{url}

\newtheorem{theorem}{Theorem} 

\newtheorem{proposition}[theorem]{Proposition}
\newtheorem{corollary}[theorem]{Corollary}

\renewcommand{\d}{\mathrm{d}}
\newcommand{\X}{\mathcal{X}}
\newcommand{\B}{\mathfrak{B}}
\newcommand{\Expect}[1]{ \mathrm{E} \! \left[ #1 \right] }
\newcommand{\T}{^{\top}}

\newcommand{\referenceMeasure}{\mu}

\newcommand{\locallyCompactHausdorff}{lcH}
\newcommand{\locallyCompactSecondCountableHausdorff}{lcscH}
\newcommand{\completelySeparableMetric}{CSM}

\newcommand{\DiracTestSeq}[2]{f_{#2,#1}}

\begin{document}

\title{Revisiting Functional Derivatives\\ in Multi-object Tracking }

\author{Jan~Krejčí,~
    Ondřej~Straka,~
    Petr~Girg,~
    and~Jiří~Benedikt
\thanks{J. Krejčí and O. Straka are with the Department of Cybernetics, University of West Bohemia in Pilsen, Pilsen, Czech Republic, e-mail: \{jkrejci / straka30\}@kky.zcu.cz.}

\thanks{P. Girg and J. Benedikt are with the Department
of Mathematics, University of West Bohemia in Pilsen, Pilsen, Czech Republic,
e-mail: \{pgirg / benedikt\}@kma.zcu.cz.}

\thanks{This research was partially supported by the European Union under the project ROBOPROX (reg. no. CZ.02.01.01/00/22\_008/0004590).}
}
%

\markboth{...}
{Shell \MakeLowercase{\textit{et al.}}: Bare Demo of IEEEtran.cls for IEEE Journals}
%

\maketitle

\begin{abstract}
	Probability generating functionals (PGFLs) are efficient and powerful tools for tracking independent objects in clutter.
	It was shown that PGFLs could be used for the elegant derivation of practical multi-object tracking algorithms, e.g., the probability hypothesis density (PHD) filter. 
	However, derivations using PGFLs use the so-called functional derivatives whose definitions usually appear too complicated or heuristic, involving Dirac delta ``functions''. 
	This paper begins by comparing different definitions of functional derivatives and exploring their relationships and implications for practical applications. It then proposes a rigorous definition of the functional derivative, utilizing straightforward yet precise mathematics for clarity. Key properties of the functional derivative are revealed and discussed.
\end{abstract}



\IEEEpeerreviewmaketitle

\section{Introduction}
\IEEEPARstart{M}{ulti-object} tracking (MOT) deals with estimating unknown states of a varying number of objects from noisy measurements~\cite{MTT-Survey:2015}.
MOT algorithms have to deal with missing and false measurements, and unknown correspondence between objects and measurements.
Such problems appear, e.g., in radar and video surveillance, autonomous driving, robotics, seafaring, aerospace, and many others~\cite{BlackmanPopoli:1999,Bar-Shalom-et.al:2011,Streit-PPP:2010,MOT-16:2016}.

The first solutions to MOT took a bottom-up approach, relying on Kalman filtering and data association techniques~\cite{Morefield:1977,Reid:1978, Bar-Shalom-et.al:2011, BlackmanPopoli:1999}.
Although top-down approaches have been used for MOT problems much earlier in the Soviet Union~\cite{ClarkNarykovStreit-StochasticFlowsPrimer:2023}, in the Western literature, they did not appear until the 1990s~\cite{Mahler:1997}.
The approach of \cite{Mahler:1997} relies on random entities called the \emph{random finite sets} (RFS) and a toolbox for their treatment called the \emph{finite set statistics} (FISST).
RFS models represent both states and measurements, managing uncertainty in their cardinality, values and various relationships. 
FISST was specifically developed for MOT problems by employing engineering-friendly stochastic geometry ideas~\cite{Kendall:1995} using finite sets, rather than conventional probabilistic \emph{point process} modeling~\cite{Vere-Jones:2003,Vere-Jones:2008,Streit-PPP:2010} that would require using counting measures.
Later on, various relationships among the modeling approaches have been revealed in~\cite{Vo-Singh-Bayes:2004,Vo-Singh-Doucet_Sequential-RFS:2005,Mahler-Book:2007,Mahler-Book:2014,MoriChongChang:ThreeFormalism:2016}.
However, the \emph{application} of RFS/FISST to the MOT problems \emph{itself} may involve problematic steps that require further justification.
This paper addresses the insufficiently rigorous and thus problematic use of the Dirac delta ``functions'' within derivations of MOT algorithms.
Dirac deltas appear in MOT when a special functional, called the \textit{probability generating functional} (PGFL), is used and differentiated.

A PGFL can be understood as a generalization of a \textit{probability generating function} for random variables to RFSs.
PGFLs have powerful properties, and their use usually leads to a more streamlined derivation of MOT algorithms \cite{Williams-MarginalPMBM:2015,Streit-Pointilist:2015,Mahler-integralTransform:2016,Streit-Book:2021}.
However, due to the Bayesian approach adopted in MOT, the practical use of PGFLs requires obtaining a \emph{probability density function} (PDF) corresponding to the measurement process from a PGFL.
This is done by taking specific derivatives of the PGFL, usually referred to as the \emph{functional derivatives}.
A functional derivative can be considered a generalization of the \emph{gradient} for functions of infinitely many variables. 
Heuristically, it can be defined as a directional derivative in the direction of the Dirac delta~\cite{Mahler-Statistic102:2013}.
Since Dirac deltas cannot be attributed to as proper functions \cite{Amaku-EtAL:ProblemsWithDiracDelta-One:2021}, a~need for mathematically sound definition arises.

From the historical perspective, functional derivatives were commonly taken for granted in many fields~\cite{Ryder:1996-QuantumFieldTheory,Greiner-Reinhardt:FieldQuantization:1996}, \cite[pp.~204--205]{DifferentialEntropyHandbook:2013}, and little care was taken for their rigorous justification given a problem at hand \cite{Mahler:MultitargetMomentsAndAppToMTT:2001}.
The most common approach followed the calculus of variations, where functional derivatives appear as integrands~\cite[Appendix~A]{Parr-Yang:DensityFunctionalTheory:1989}.
Such an approach can be justified by interpreting the integrands as density functions \emph{almost everywhere} \cite[Appendix C]{Mahler-Book:2014},\cite{Mahler_GeneralizedPHD:2012,SchlangenDelandeHoussineauClark:SO-PHD:2018}.
However, the derivatives must be defined point-wise, since the measurement density function (i.e., multi-object likelihood) needs to be defined at any particular zero-probability measurement set received from a sensor.
This is for applying Bayes' rule correctly~\cite[p.~7]{Mahler-Statistic103:2019}.
Moreover, to be of practical use, the derivative must be generally applicable to PGFLs (and its ``parts'') and obey differentiation rules, such as the chain rule \cite{clark2012faa,Mahler_GeneralizedPHD:2012} in the case of MOT.

Apparently, by drawing inspiration from the quantum field theory~\cite[p.~173]{Ryder:1996-QuantumFieldTheory}, functional derivatives were introduced to the western MOT literature by Mahler in~\cite{Mahler-PHD:2003} to derive the well-known \emph{probability hypothesis density} (PHD) filter. Therein, the derivatives were established point-wise as heuristic directional derivatives.
In~\cite{Mahler:1997}, they were defined via the so\discretionary{-}{-}{-}called \emph{set derivatives}, which was later promoted as being a rigorous approach~\cite[Appendix C.3]{Mahler-Book:2014}. 
However, set derivatives appear mathematically involved and less established than the classic approach described before.
Consequently, the theory of distributions was used in~\cite{Streit_FunctionalDerivatives:2015} as an attempt to make the directional derivative approach rigorous.
However, as will be shown in this paper, all the currently known approaches, including the use of the theory of distributions, appear insufficient in general. 

Technical details regarding mathematical rigor, such as the validity regions of PGFLs and subsequently of the derivatives, are usually omitted or simplified in the MOT literature \cite{Vo-Singh-Bayes:2004,Mahler-PHD:2003,Mahler-Book:2014,Streit-Book:2021}.
Moreover, the existence of various definitions of the functional derivative makes the overall theory less clear.
Therefore, these definitions are reviewed and compared in this paper, emphasizing both mathematical rigor and simplicity.
Practical usage and implications of the definitions are summarized, and a convenient definition of functional derivative is proposed with minimum drawbacks.

The main objectives of this paper are: 
\begin{itemize}
    \item To review the current functional derivative definitions with a particular focus on the MOT problem and to show theoretical problems in the definitions.
    \item To provide a rigorous and practically useful re-definition of functional derivative and give its basic properties.
\end{itemize}

The paper is organized as follows. 
Section~\ref{sec:preliminaries} briefly summarizes the background on probabilistic modeling used in MOT.
Existing definitions of functional derivatives are reviewed and compared in Section~\ref{sec:definitions}.
A suitable and mathematically valid definition is proposed and discussed in Section~\ref{sec:proposed}.
Concluding remarks are given in Section~\ref{sec:conclusion} while mathematical proofs are given in appendices.
Further mathematical details are reviewed in Supplementary Materials.

\section{Background on RFS Modeling}\label{sec:preliminaries}
This section summarizes the background of modeling MOT problems in a top-down manner, in which objects and measurements are modeled as finite point patterns.
Such patterns can be represented using finite sets, whose randomization leads to the RFS theory and can be endowed with the FISST notions.
Other representations can be applied to MOT problems having a strong correspondence among them~\cite[Proposition~5.3.II]{Vere-Jones:2003}\cite[p.~525]{Kallenberg-book:2002}\cite{Vo-Singh-Bayes:2004,Vo-Singh-Doucet_Sequential-RFS:2005,Mahler-Book:2007,Mahler-Book:2014,MoriChongChang:ThreeFormalism:2016}.
The RFS formalism is thus adopted throughout this paper, having the following facts in mind: 
\begin{itemize}
    \item It is well-established in the MOT literature.
    \item It has a simple notation compared to other approaches.
    \item It can unify all levels of information fusion with some effort~\cite{Mahler:1997,Mahler-Book:2007,Mahler-Book:2014}.
    \item It has a simple geometric interpretation as sets of points.
    \item An RFS is essentially a \emph{simple finite point process} \cite{MoriChongChang:ThreeFormalism:2016}.
    \item RFS elements cannot be independent and identically distributed by the nature of forming a set~\cite{Streit-HowILearned:2017}, which can be alleviated by accounting for multisets (i.e., model multiple identical elements) with some effort~\cite[Sec.~5.2.3]{Mahler:1997}\cite[Appendix~E]{Mahler-Book:2007}.
\end{itemize}
Correspondences with point processes are made when appropriate, see also Supplementary Materials (Supplement~\ref{appendix:RFS_notes}).

\subsection{The Base Space}\label{sec:Background:TheBaseSpace}
The space $\X$, in which an RFS elements lie, is referred to as the \emph{base space}.
The random set treatment generally requires $\X$ to be \emph{locally compact}, \emph{second countable}, and \emph{Hausdorff} (\locallyCompactSecondCountableHausdorff{}) topological space~\cite{Mahler:1997}. 
For convenience, the most relevant facts about \locallyCompactSecondCountableHausdorff{} spaces are given in Supplement~\ref{appendix:topology}.
Note that $\X {=} \mathbb{R}^d$ (endowed with its usual topology) is a common choice for modeling objects that are described with only several kinematic variables, such as position and velocity.
However, by assuming $\X$ to be \locallyCompactSecondCountableHausdorff{} in general, it is possible to account for broader field of applications, where objects may have distinct marks or varying dimensions~\cite[Sec. 2.2.2]{Mahler-Book:2014}.
For instance, denoting disjoint union with $\uplus$, one could use $\X {=} \mathbb{R}^{n_1} {\uplus} \mathbb{R}^{n_2}$ in a multiple model approach to tracking maneuvering objects, or when clutter generators are modeled \cite[Chapter~18]{Mahler-Book:2014}.

It is assumed that $\X$ is endowed with a measure-theoretic integration concept.
Let $(\X,\B,\referenceMeasure)$ be a measure space with $\B$ being the Borel $\sigma$-algebra and $\referenceMeasure$ a reference Radon measure, see Supplement~\ref{appendix:Radon_Measures}.
If, for instance, $\X {=} \mathbb{R}^d$, then it is usual to choose $\referenceMeasure$ as the $d$-dimensional Lebesgue measure\footnote{
    More precisely, the restriction of the Lebesgue measure to the standard Borel $\sigma$-algebra on $\mathbb{R}^d$.
}.
For simplicity, integration w.r.t.~the reference Radon measure $\referenceMeasure$ is denoted as
\begin{align}
    \int_{\X} f(\mathbf{x}) \referenceMeasure (\d \mathbf{x}) = \int_{\X} f(\mathbf{x}) \d \mathbf{x} \, .
    \label{eq:integral-wrt-reference-measure-notation}
\end{align}

If $f$ is a probability density on $\X$, the integral~\eqref{eq:integral-wrt-reference-measure-notation} is assumed to be one, which is the usual property one would expect if $\referenceMeasure$ was the Lebesgue measure.
If $\referenceMeasure$ is translation invariant\footnote{
    Such a measure cannot be defined on any \locallyCompactSecondCountableHausdorff{} space; see \cite[Chapter~11]{Folland:1999}.
    For $\X {=} \mathbb{R}^d$, such $\referenceMeasure$ is only the Lebesgue measure up to a scaling factor.
    For more general problems encountered in practice such as for $\X {=} \mathbb{R}^{n_1} {\uplus} \mathbb{R}^{n_2}$, the Lebesgue measure can be generalized straightforwardly.
}, it is convenient to introduce units of measurement on $\X$.
Denote the unit of measurement in $\X$ as $u$, i.e., $u=\referenceMeasure(U)$ with $U\in\B$ being a chosen ``unit set'' of the space $\X$.

\subsection{Brief Introduction to RFSs}
An RFS $\Xi$ is a random subset of $\X$ with a finite number of elements.
That is, both the cardinality $|\Xi|$ and the elements $\mathbf{x} \in \Xi$ are random.
The realizations of $\Xi$ will be further denoted with $X=\{\mathbf{x}_1, \dots, \mathbf{x}_{n}\}$ for notational simplicity, i.e., random elements on $\X$ and their realizations are both denoted with lower-case bold-face symbols such as~$\mathbf{x}$.

Any realization $X$ can be seen as an element of a hyperspace comprised of all finite subsets of $\X$, denoted with $\mathcal{F}(\X)$.
Omitting details\footnote{
    To introduce the probability measure of $\Xi$ correctly, the space $\mathcal{F}(\X)$ has to be turned into a probability space, and $C$ should be taken as a Borel measurable subset of $\mathcal{F}(\X)$.
    This can be done by adopting the Mathéron ``hit-or-miss'' topology on $\mathcal{F}(\X)$ and inducing a $\sigma$-algebra.
    The details can be found, e.g., in \cite{Mahler:1997,Vo-Singh-Doucet_Sequential-RFS:2005,Vihola:PhD:2004}\cite[Appendix~F]{Mahler-Book:2007}.
} not needed for this paper, the probability measure corresponding to $\Xi$ is denoted with
\begin{align}
    P_{\Xi}(C) \triangleq \Pr[ \Xi \in C ], \label{eq:probability_measure_RFS}
\end{align}
where $C$ is any measurable subset of $\mathcal{F}(\X)$.
Such an abstract entity provides little tractability.
The entities introduced in the following shall be more handy in describing RFSs.

\subsection{Belief Mass Function}
The \emph{belief mass function} (BMF) $\beta_\Xi(S)$ is a restriction of the measure $P_{\Xi}(C)$~\eqref{eq:probability_measure_RFS} to sets of the form $C{=}\uplus_{n=0}^{+\infty} S^{(n)}$  parametrized with a closed subset $S\subseteq \X$, where $S^{(n)}\subseteq \mathcal{F}(\X)$ is the set of (measurable) subsets of $S$ with cardinality $n$ and $S^{0} {=} \{\emptyset\}$ by convention, i.e., ~\cite{Vo-Singh-Bayes:2004}
\begin{align}\label{eq:BMF_RFS}
    \beta_{\Xi}(S) \triangleq P_{\Xi}(\uplus_{n=0}^{+\infty} S^{(n)}) = \Pr[ \Xi \subseteq S ].
\end{align}
That is, a BMF is the probability that the entire realization of the RFS lies in the set $S$, and it is thus not additive in $S$, i.e., it is not a measure.
Surprisingly, by the Choquet theorem known from the general random set theory, a BMF characterizes an RFS completely~\cite[p.~713]{Mahler-Book:2007}, and therefore it can be used to define a density function corresponding to the RFS.

\subsection{Density Function}\label{subsec:density_function}
A density function of an RFS is a real, non-negative function whose arguments are finite sets, such as the set of objects or measurements.
They are essential for MOT problems because of the Bayes rule, where the likelihood function has to be evaluated at a given measurement set.

In conventional probability theory, a \emph{probability density function} is defined as the Radon--Nikodým derivative of the corresponding probability measure.
This is adopted in the point process approach, where the derivative is studied, for example, w.r.t.~the unit-rate Poisson point process~\cite[Sec.~10.4]{Vere-Jones:2008}\cite{Vo-Singh-Bayes:2004}.
Alternatively, one can describe the point process with a sequence of the so-called Janossy measures $\{J_n\}_{n\geq0}$, each being a measure on $\X^n$, see Supplement~\ref{appendix:RFS_notes}. 
Assuming $J_n$ is absolutely continous w.r.t.~the $n$-fold product of the reference measure $\referenceMeasure$, denoted with $\referenceMeasure^n$, the~Radon--Nikodým derivative exists as
\begin{align}
    j_n(\mathbf{x}_1,\dots,\mathbf{x}_n) \triangleq \frac{\d J_n}{\d \referenceMeasure^n} (\mathbf{x}_1,\dots,\mathbf{x}_n) \, , \label{eq:Janossy:density}
\end{align}
and is called the $n$-th order Janossy density~\cite[Sec.~10.4]{Vere-Jones:2008}.
In this case, the points form a \emph{set} with probability one and the Janossy densities~\eqref{eq:Janossy:density} vanish on the diagonal, i.e., whenever $\mathbf{x}_i {=} \mathbf{x}_j$ for any $i {\neq} j$.
The Radon--Nikodým derivative, however, defines the Janossy densities only $\referenceMeasure$-almost everywhere \mbox{($\referenceMeasure$-ae)}.
As a result, the likelihood function defined this way could evaluate to any real number for a given measurement set without contradictions.
This can be addressed, e.g., via the Lebesgue differentiation theorem~\cite[Section~3.4]{Folland:1999}.
In the context of RFSs, this has been done as follows.

The FISST approach~\cite{Mahler:1997} constructs the density from the BMF $\beta_\Xi(S)$ \eqref{eq:BMF_RFS} via the \emph{set derivative} as
\begin{align}\label{eq:PDF_from_BMF}
    p_\Xi(X) \triangleq \left[ \frac{\delta}{\delta X} \beta_\Xi (S) \right]_{S=\emptyset},
\end{align}
where the set derivative denoted with ${\delta} / {\delta X}$ is iterative, and it is given in Supplement~\ref{appendix:set_derivatives} for completeness.
If the cardinality of an RFS is bounded\footnote{
    In~\cite{Mahler:1997}, the function $p_\Xi(X)$~\eqref{eq:PDF_from_BMF} must vanish identically for all sufficiently large cardinalities $|X|$.
    It follows that there is no ``Poisson RFS'' under the formulation of~\cite{Mahler:1997} unless it is an approximation, see~\cite[pp.~223,~235]{Mahler:1997}.
}, it was shown in~\cite[pp.~156--168]{Mahler:1997}, that the density $p_\Xi(X)$~\eqref{eq:PDF_from_BMF} exists under appropriate notion of absolute continuity of $\beta(S)$~\eqref{eq:BMF_RFS}.
Nevertheless, these conditions can be extended to the unbounded RFS framework~\cite{Mahler:1997,Vo-Singh-Doucet_Sequential-RFS:2005,Mahler-Book:2007} adopted nowadays, see the Licentiate thesis~\cite{Vihola:PhD:2004}.
If both the densities~\eqref{eq:Janossy:density} and~\eqref{eq:PDF_from_BMF} exist, the density $p_\Xi(X)$~\eqref{eq:PDF_from_BMF}, when evaluated at the set $X=\{\mathbf{x}_1,\dots,\mathbf{x}_n\}$, is $\referenceMeasure$-ae equal to the $n$-th order Janossy density~\cite[Eq.~(38)]{Mahler-PHD:2003},
\begin{align}
    p_{\Xi}(\{ \mathbf{x}_1,\dots,\mathbf{x}_n \}) = j_n(\mathbf{x}_1,\dots,\mathbf{x}_n) \, .
    \label{eq:densities:FISST-Janossy}
\end{align}
This establishes the key practical connection between the RFS and point process approaches.
In fact, all the mentioned approaches to the definition of an RFS density are \emph{equivalent} in the sense that the resulting functions can be expressed ($\referenceMeasure$-ae) in terms of each other \cite{MoriChongChang:ThreeFormalism:2016,Mahler-PHD:2003}, see Supplement~\ref{appendix:RFS_notes}.

Note that the unit of measurement of $p_\Xi(X)$~\eqref{eq:densities:FISST-Janossy} is $u^{-|X|}$, i.e., it varies with the cardinality of $X$~\cite[p.~163]{Mahler:1997}.

\subsection{Set Integral}
In order to construct, e.g., marginal densities or moments for RFSs, a notion of integration is needed.
One can either stay in the hyperspace $\mathcal{F}(\X)$, or define integrals straight in the space $\X$ and make use of the Janossy densities~\cite[p.~14]{SUMMER_FISST:2013}.
The former approach is quite involved, but classic properties of measure-theoretic integrals persist.
The latter approach is taken, for example, by the \emph{set integral}; it is more intuitive on the one hand, but the integral is not additive in the domain of integration on the other hand.

The set integral of a finite-set function $f$, i.e., a function whose argument is a finite set, is defined as
\begin{align}
    \int_{\mathcal{X}} \! f(X) \delta X \triangleq \! \sum_{n=0}^{+\infty} \frac{1}{n!} \underbrace{ \int_{\X}\!\!\cdots\!\int_{\X} }_{n \text {-times}} \! f(\{\mathbf{x}_1,\dots,\mathbf{x}_n\}) \d\mathbf{x}_1\cdots \d\mathbf{x}_n , \notag\\[-0.6cm]
    \label{eq:def_set_Integral}
\end{align}
and it is the ``anti-derivative'' corresponding to~\eqref{eq:PDF_from_BMF}~\cite[p.~159]{Mahler:1997}.
Therefore, $\int_\X p_\Xi(X) \delta X = 1$.
The first term of~\eqref{eq:def_set_Integral} corresponding to $n=0$ is a constant $f(\emptyset)$.

\subsection{Probability Generating Functional}
A density function is a convenient tool for describing RFSs due to its resemblance of the single-object case.
Working with densities, however, might get occasionally complicated, calling for different tools.
One such tool is the \emph{probability generating functional} (PGFL), defined as follows.

Let $\Gamma(\X)$ be the class of complex-valued Borel measurable functions $h:\X\rightarrow\mathbb{C}$ such that $|h(\mathbf{x})| {\leq} 1, \forall \mathbf{x} {\in} \X$.
Note that any continuous or piecewise constant (i.e., \emph{simple}) function is Borel measurable~\cite[Sec.~2.1]{Folland:1999}.
Furthermore, consider the exponential notation 
\begin{align}
    h^X = \begin{cases}
        1 & \text{if } X = \emptyset, \\
        \prod_{\mathbf{x}\in X} h(\mathbf{x}) & \text{if } X = \{\mathbf{x}_1,\dots,\mathbf{x}_n\} \, .
    \end{cases} \label{eq:def_hX}
\end{align}
The PGFL $G_\Xi:\Gamma(\X)\rightarrow \mathbb{C}$ of an RFS $\Xi$ can then be defined as the expectation\footnote{
    The PGFL can be defined over a narrower class of certain real-valued functions as a special case~\cite[Sec.~9.4]{Vere-Jones:2008}, but the advantages of doing so are not needed in this paper.
}~\cite[Sec.~5.5]{Vere-Jones:2003}, \cite{Mahler-PHD:2000}
\begin{align}
    G_{\Xi}[h] &\triangleq \Expect{ h^\Xi } = \int_{\mathcal{X}} h^{X} p_{\Xi}(X) \delta X \, .
    \label{eq:def_PGFL} 
\end{align}
Notice that $G_\Xi[0] {=} p_\Xi(\emptyset) {=} \Pr[\Xi{=}\emptyset]$, and $G_\Xi[1] {=} 1$. 
The most powerful advantage of using PGFLs over density functions emerges when dealing with a union, also known as superposition, of \emph{independent} RFSs.
It can be straightforwardly seen that the PGFL corresponding to a union of independent RFSs is given by the product of their PGFLs~\cite[p.~63]{Vere-Jones:2008},\cite[Sec.~4.2.4]{Mahler-Book:2014}.
A corresponding density function is given by a considerably more involved \emph{convolution formula}~\cite[Sec.~4.2.3]{Mahler-Book:2014}. 
The product-property applies to the BMFs as well, since $\beta_\Xi(S) = G_\Xi[\mathbf{1}_S]$, where $\mathbf{1}_S(\mathbf{x})$ is the indicator function of the closed set $S\subseteq\X$~\cite{Mahler-PHD:2003}.
Nevertheless, PFGLs are defined over functions that can be multiplied and added unlike sets which form the domain of the BMF.

PGFLs are thus powerful tools to ease both the treatment of RFSs and the notation involved.
On the other hand, one might occasionally need to derive the density function from the corresponding PGFL, e.g., when the likelihood function must be evaluated at a given measurement set.
To do so, functional derivatives are used.

\subsection{Functional Derivatives in MOT}\label{sec:functional_derivatives_in_MTT}
This section provides only the heuristic definition of the functional derivatives as the directional derivative, i.e.,
\begin{align}
    \frac{\delta G_{\Xi}}{\delta \mathbf{x}}[h] \triangleq \frac{\partial G_{\Xi}}{\partial \delta_{\mathbf{x}} } [h] = \lim_{\epsilon \rightarrow 0} \frac{G_{\Xi}[h+\epsilon \delta_{\mathbf{x}}] - G_{\Xi}[h] }{\epsilon} \, , \label{eq:def_functional_derivative}
\end{align}
where $\delta_{\mathbf{x}}(\mathbf{x}')$ is the Dirac delta ``function''.
Obviously, \eqref{eq:def_functional_derivative} is ill-defined as $(h {+} \epsilon\delta_{\mathbf{x}})$ is certainly not an element of the class $\Gamma(\X)$ as $\delta_{\mathbf{x}}$ is not a function, see Supplement~\ref{appendix:Dirac_deltas}.
Iterated derivatives w.r.t. a finite set $X {=} \{ \mathbf{x}_1, \dots, \mathbf{x}_n \} \subset \X$, $n \in \mathbb N {\cup} \{0\}$, are formally defined as 
\begin{subequations}\label{eq:def_functional_derivative_iterated}
\begin{align}
    \frac{\delta G_{\Xi}}{\delta \emptyset}[h] &= G_{\Xi} [h] \,, \label{eq:def_functional_derivative_iterated_1} \\
    \frac{\delta G_{\Xi}}{\delta \{ \mathbf{x}_1, \dots, \mathbf{x}_n \} }[h] &= \frac{\delta}{\delta \mathbf{x}_n } \frac{\delta G_{\Xi}}{\delta \{ \mathbf{x}_1, \dots, \mathbf{x}_{n-1} \} }[h] \label{eq:def_functional_derivative_iterated_2} \, . 
\end{align}
\end{subequations}

Formally, functional derivatives obey many analytic differentiation rules that are significantly reminiscent of the classic calculus with functions \cite[Sec.~3.5]{Mahler-Book:2014}.
Although similar rules apply to set derivatives of BMFs, working with PGFLs appears to be more convenient and more generally accepted outside the theory developed for solving MOT problems~\cite{Streit-Book:2021}.

Functional derivatives can be used for finding extremes of functionals in optimization~\cite{Gelfand-Fomin:Variations:1963,Volterra:funcionalCalcullus:1930,GradientBoosting:1999,DensityFunctionalTheory:2011}. 
In physics~\cite{Ryder:1996-QuantumFieldTheory,Greiner-Reinhardt:FieldQuantization:1996} or statistics~\cite[pp.~204--205]{DifferentialEntropyHandbook:2013}, these methods are commonly mathematically imprecise, similarly to their daily use in MOT.
In MOT, functional derivatives are mostly used to \emph{recover} a density function from a known PGFL expression or to compute moments corresponding to the RFS as follows.

\subsubsection{PGFL $\rightarrow$ density function}
The density function $p_\Xi(X)$~\eqref{eq:densities:FISST-Janossy} can be computed from the PGFL $G_\Xi[h]$~\eqref{eq:def_PGFL} using functional derivatives as~\cite[Sec.~3.5.1]{Mahler-Book:2014}
\begin{align}
    p_\Xi(X) &= \left[ \frac{\delta }{\delta X} G_\Xi[h] \right]_{h=0} = \frac{\delta G_\Xi }{\delta X} [0] \, .
    \label{eq:PGFL_to_PDF}
\end{align}

\subsubsection{PGFL $\rightarrow$ factorial moment}
The \emph{factorial moment} $D_\Xi(X)$ is an RFS analogue to raw statistical moment~\cite[Sec.~2.1.5]{Mahler:MultitargetMomentsAndAppToMTT:2001}.
If it exists, it can be computed as~\cite[Sec.~4.2.9]{Mahler-Book:2014}
\begin{align}
    D_\Xi(X) &= \left[ \frac{\delta }{\delta X} G_\Xi[h] \right]_{h=1} = \frac{\delta G_\Xi }{\delta X} [1] \, . \label{eq:PGFL_to_MomentDensities}
\end{align}
If $X=\{ \mathbf{x} \}$, the function $D_\Xi(\{ \mathbf{x} \})$ is known as the PHD~\cite[Sec.~4.2.8]{Mahler-Book:2014} or \emph{intensity}, and it is central to the well-known PHD~\cite{Mahler-PHD:2003}, CPHD~\cite{Mahler-CPHD:2007} or intensity filters~\cite{Streit-PPP:2010,Streit-deriv:2008}.

Other applications of functional derivatives in MOT include the computation of factorial cumulant densities~\cite{clark2012faa,ClarkDeMelo-2ndOrdCumulantFilter:2018}, or Palm conditioning~\cite{Streit-Bozdogan-Efe_Palm:2013}.

\subsection{Problem Statement}\label{sec:problem-statement}
The central problem of this paper is the \emph{definition} of functional derivatives, including their higher-order versions~\eqref{eq:def_functional_derivative_iterated} for functionals appearing in MOT.
If the density function and moments exist, Eq.~\eqref{eq:PGFL_to_PDF} and~\eqref{eq:PGFL_to_MomentDensities}
must hold, i.e., the derivatives shall be possible to evaluate at $h {\equiv} 0$ and $h {\equiv} 1$.
Properties of the RFS density that guarantee the existence of the derivative should also be given, so that engineers need not check the aforementioned existence.
Regarding the directions $\delta_{\mathbf{x}_1},\dots,\delta_{\mathbf{x}_n}$, the derivatives shall be defined for any disjoint elements $\mathbf{x}_1,\dots,\mathbf{x}_n$ forming the set $X$, i.e., point-wise ideally everywhere in $\X$. 
Moreover, the derivative framework should obey formal differentiation rules
and thus be applicable not only to PGFLs themselves but also provide means to work with them effectively.
Note that the chain rule is key, so that if the formal rules~\cite[Section~3.5]{Mahler-Book:2014} hold, they can be used iteratively.
That is, the definition should allow for the practical use of the derivatives.
The main goal of this paper is to seek a definition of functional derivatives that is simple, mathematically rigorous, general, and thus reliable and useful.

\section{State of the Art in Functional Derivatives: Definitions and Analysis}\label{sec:definitions}
In this section, existing definitions of functional derivatives are stated as presented in the literature and reviewed. 
Assumptions that are usually missing in the MOT literature are added and critically assessed.

It should be noted that any existing definition of the functional derivative, regardless it is rigorous or not, shall not influence its purely practical usage described, e.g., in~\cite{Mahler-Book:2014,Streit-Book:2021}.
However, using the wrong definition may lead to erroneous conclusions on the theoretical level.
In general, using speculative mathematics can be dangerous and has severe consequences~\cite{Jaffe-Quinn:TheoreticalMath:1993}. 
In the case of MOT, this could lead to accounting algorithms as Bayes-optimal in certain applications, even though they could not be derived correctly under given circumstances, resulting in being merely ad-hoc solutions.

The common starting point for all the following definitions may be the directional (variational) derivative of a functional $G[h]$ in the direction of $g$, which is formally given by 
\begin{align}
    \frac{\partial G[h]}{\partial g} &\triangleq
    \lim_{\epsilon \rightarrow 0} \frac{G[h \! + \! \epsilon \! \cdot \! g] - G[h] }{\epsilon} \, . \label{eq:directional_derivative_PFGL}
\end{align}
Properties and ranges of validity are given when dealing with a particular definition, based on standard functional analysis techniques reviewed in Supplement~\ref{appendix:derivatives_NLP}.
Under some assumptions on $G[h]$, the derivative~\eqref{eq:directional_derivative_PFGL} may be written in the form
\begin{align}
    \frac{\partial G[h]}{\partial g} = \int_{\mathcal{X}} T[h](\mathbf{x}) g(\mathbf{x}) \d\mathbf{x} \, , \label{eq:directional_derivative:with_kernel}
\end{align}
where $h \, {\mapsto} T[h](\cdot)$ is a functional transformation.
The functional transformation itself can be heuristically obtained by 
\begin{align}
    \frac{\partial G[h]}{\partial \delta_{\mathbf{x}}} = T[h](\mathbf{x})\, ,
\end{align}
which thus has the meaning of the first-order functional derivative $\frac{\delta G[h]}{\delta \mathbf{x}}$. 
The following text reviews approaches to defining the functional derivative in the MOT setting.

\subsection{Definition Bypassing Limit Procedure}\label{subsec:definition_for_PGFLS}
Formally it can be shown, that the directional derivative~\eqref{eq:directional_derivative_PFGL} of the PGFL $G_{\Xi}[h]$~\eqref{eq:def_PGFL}, can be expressed as \cite{Streit-Bozdogan-Efe_Palm:2013}, \cite[p.~12]{Mahler-Statistic103:2019}
\begin{align}
    \frac{\partial G_{\Xi}[h]}{\partial g} = \int_{\mathcal{X}} \left( \int_{\mathcal{X}} h^X p_{\Xi}(X\cup \{\mathbf{x}\}) \delta X \right) g(\mathbf{x}) \d\mathbf{x} \, , \label{eq:functional_derivative_allwaysExist}
\end{align}
and thus the functional derivative becomes
\begin{align}
    \frac{\delta G_{\Xi}[h]}{\delta \mathbf{x}} = \int_{\mathcal{X}} h^X p_{\Xi}(X\cup \{\mathbf{x}\}) \delta X \, . \label{eq:PGFL_kernel}
\end{align}
Iterated functional derivatives~\eqref{eq:def_functional_derivative_iterated_2} can be formally revealed in this fashion as well \cite{Streit-Bozdogan-Efe_Palm:2013,Streit:PGFLsPHDvsIntensity:2013}, yielding
\begin{align}
    \frac{\delta G_{\Xi}[h]}{\delta Y} = \int_{\mathcal{X}} h^X p_{\Xi}(X\cup Y) \delta X \, , \label{eq:PGFL_kernel_iterated}
\end{align}
where $Y = \{\mathbf{x}_1, \dots, \mathbf{x}_n\} \subset \X$ is a finite set.
Functional derivatives of a PGFL and the iterated versions could then be defined directly as the right-hand sides of~\eqref{eq:PGFL_kernel} and~\eqref{eq:PGFL_kernel_iterated}, respectively.
However, such a definition is not practical and does not appear in the literature.

\subsubsection*{Analysis of the definition}
this approach is specific to PGFLs and differentiation rules are unavailable.
Therefore, using such definition would force the user to prove that a particular result derived using differentiation rules indeed has the desired form, which is impractical.
On the other hand, such a definition would be valid for any $h {\in} \Gamma(\X)$ while it would not pose any further requirements on the density $p_{\Xi}(X)$.
In fact, the PGFL could be defined using the Janossy \emph{measures}~\cite[p.~145]{Vere-Jones:2003} directly for~\eqref{eq:PGFL_kernel}-\eqref{eq:PGFL_kernel_iterated}, even if they does not have densities w.r.t.~the reference measure $\referenceMeasure$, e.g., if $\Xi$ contains multiplicities.
Nevertheless, if the density $p_{\Xi}(X)$ is not defined point-wise, the resulting derivative is neither, i.e. for a particular $\mathbf{x}$ in the case of~\eqref{eq:PGFL_kernel} or $Y$ in the case of~\eqref{eq:PGFL_kernel_iterated}.
To define it point-wise, $p_{\Xi}(X)$ could be strictly taken as the FISST density function~\eqref{eq:PDF_from_BMF} that is defined point-wise. 
\\[-0.2cm]

For a more practical definition, the functional transformation of~\eqref{eq:directional_derivative:with_kernel} can be constructed, regardless its specific form~\eqref{eq:PGFL_kernel} and~\eqref{eq:PGFL_kernel_iterated} for PGFLs, as described in the following Subsection.

\subsection{Definition Almost Everywhere}\label{subsec:definition_ae}
This construction follows~\cite{Mahler_GeneralizedPHD:2012} and uses~\cite[Appendix~C1]{Mahler-Book:2014}.
Assume that a space $\mathrm{F}(\X)$ of functions is a Banach space, i.e., a complete normed vector space.
Although not specified in~\cite{Mahler_GeneralizedPHD:2012}, we may assume that $\Gamma(\X) {\subset}\, \mathrm{F}(\X)$.
Consider a functional $G[h]$ on $\mathrm{F}(\X)$ such that
\begin{align}
    G[h+q]=G[h] \, , \label{eq:property_of_nice_functionals}
\end{align}
for any function $q$ whose support\footnote{
    According to~\cite[p.~1045]{Mahler-Book:2014}, the support of function $q$ is defined as $\mathrm{Supp}(q) \triangleq \{ \mathbf{x}\in\X: q(\mathbf{x})\neq 0 \}$, cf. Eq.~\eqref{app:support-topology} from Supplement~\ref{appendix:topology}.
} has $\referenceMeasure$-measure zero, i.e., $q$~vanishes almost everywhere. 
This construction assumes that the directional derivative of such functional is moreover a Gâteaux derivative, i.e, it is linear and continuous w.r.t. the direction $g$, see Supplement~\ref{appendix:derivatives_NLP}.
With this, the derivative~\eqref{eq:directional_derivative_PFGL} in the direction of the indicator function $g(\mathbf{x})=\mathbf{1}_{S}(\mathbf{x})$ with $S\subset \X$ being measurable, defines a $\sigma$-finite measure as
\begin{align}
    \phi_h(S) = \tfrac{\partial G}{\partial \mathbf{1}_S} [h] \, . \label{eq:measure_phi}
\end{align}
Thanks to~\eqref{eq:property_of_nice_functionals}, the measure $\phi_h(S)$~\eqref{eq:measure_phi} is absolutely continuous w.r.t.~$\referenceMeasure$ and thus the Radon\discretionary{--}{-}{--}Nikodým theorem can be used to define its density $\frac{\d \phi_h}{\d \mathbf{x}}(\mathbf{x}) {=} \frac{\d \phi_h}{\d \referenceMeasure}(\mathbf{x})$ $\referenceMeasure$-ae, and
\begin{align}
    \frac{\partial G[h]}{\partial \mathbf{1}_{S}} \! = \!
    \int_S \! \d \phi_h \! =\! \int_S \! \tfrac{\d \phi_h}{\d \mathbf{x}}(\mathbf{x}) \d\mathbf{x} \!=\!
    \int_{\mathcal{X}} \! \tfrac{\d \phi_h}{\d \mathbf{x}}(\mathbf{x}) \mathbf{1}_{S}(\mathbf{x}) \d\mathbf{x}
    \, .
\end{align}
Thanks to the theorem known as \emph{approximation by simple functions}~\cite[2.10~Thm., p.~47]{Folland:1999}, any direction $g$ can be written as a limit of simple functions and thus~\cite[Appendix~C1]{Mahler-Book:2014}
\begin{align}
    \frac{\partial G[h]}{\partial g} =
    \int_{\mathcal{X}} \frac{\d \phi_h}{\d \mathbf{x}}(\mathbf{x}) g(\mathbf{x}) \d\mathbf{x} \, ,
\end{align}
which means that $\frac{\d \phi_h}{\d \mathbf{x}}(\mathbf{x})$ can be used as a definition of the functional derivative $\frac{\delta G}{\delta \mathbf{x}}[h]$ almost everywhere.
The iterated version is then defined by~\eqref{eq:def_functional_derivative_iterated_2}.

\subsubsection*{Analysis of the definition}
if the Gâteaux derivative is moreover continuous in $h$, or it is a Fréchet derivative, it can be shown to yield the chain rule, see Supplement~\ref{appendix:derivatives_NLP}.
If such a definition is used with a PGFL~\eqref{eq:def_PGFL}, the assumption~\eqref{eq:property_of_nice_functionals} only means that the density $p_{\Xi}(X)$ must exist, i.e., that the Janossy measures are absolutely continuous w.r.t.~$\referenceMeasure$. 
The very assumption that the derivative is Gâteaux, however, is violated by the fact that PGFLs are defined on the space $\Gamma(\X)$, which is not a vector space due to the constraint $|h(\mathbf{x})| {\leq} 1, \forall \mathbf{x} {\in} \X$.
That is, the functional derivative defined this way cannot be a Gateaux derivative, and thus it does not support the chain rule in general.
The existence of the derivative must be checked as no specific conditions for $p_{\Xi}(X)$~\eqref{eq:densities:FISST-Janossy} are provided.
The derivative is moreover not defined point-wise.
\\[-0.2cm]

Note that some authors seem to be satisfied with this definition, assuming it is a Chain differential (a concept in between Gâteaux and Fréchet differentials, yielding the chain rule, see Supplement~\ref{appendix:derivatives_NLP})~\cite{clark2012faa,ClarkDeMelo-2ndOrdCumulantFilter:2018}.
As discussed before, a $\referenceMeasure$-ae definition of the functional derivative leads to ambiguities.
To define it point-wise, one can \emph{construct} the Radon--Nikodým derivative of the measure $\phi_h(S)$~\eqref{eq:measure_phi} using a Lebesgue differentiation theorem, as described in the following subsection.

\subsection{FISST Definition via Set Derivative}\label{subsec:definition_set_derivatives}
This construction follows~\cite{Mahler-Statistic102:2013}, \cite[Appendix~C.3]{Mahler-Book:2014} and is referred to as the FISST definition.
Therein, the functional derivative is defined point-wise as
\begin{align}
    \frac{\delta G}{\delta \mathbf{x}} [h]
    &= \left[ \frac{\delta }{\delta \mathbf{x}} \frac{\partial G}{\partial \mathbf{1}_S} [h] \right]_{S = \emptyset} \, , \label{eq:functional_derivative:Mahlers_definition}
\end{align}
where the outer derivative on the right-hand side is the set derivative, see Supplement~\ref{appendix:set_derivatives}.
The set derivative constructs the density of the measure $\phi_h(S)$~\eqref{eq:measure_phi} wherever the limit, i.e., the set derivative, exists.
The iterated version is given by~\eqref{eq:def_functional_derivative_iterated_2}.

\subsubsection*{Analysis of the definition}
this definition inherits the properties of the preceding definition while defining the derivative point-wise.
Although being extensively referred to in MOT, set derivatives apparently appear in the MOT (and closely related) literature only.
Some authors thus wish to provide alternative definitions~\cite{Streit_FunctionalDerivatives:2015,Streit-Book:2021,Streit-PHD_ordinary_derivatives:2014} utilizing more widely accepted techniques. 
Note that the definitions presented so far do not discuss any further conditions on the density function $p_{\Xi}(X)$~\eqref{eq:densities:FISST-Janossy} under which the derivatives are guaranteed to exist.
\\[-0.2cm]

From early approaches to the calculus of variations, the \emph{Volterra} derivative~\cite[Sec.~2.1]{Volterra:funcionalCalcullus:1930} is constructed similarly to~\eqref{eq:directional_derivative_PFGL}.
Volterra, however, used a single (multivariable) limit that \emph{shapes} the direction $g$ to \emph{mimic} the properties of the Dirac delta, that could be seen as essentially the same requirements as the limits in~\eqref{eq:functional_derivative:Mahlers_definition} apply ``together''.
In fact, Volterra defined the derivative for functionals whose arguments are functions of a single real variable~\cite[Sec.~II.1]{Volterra:funcionalCalcullus:1930}.
From this point of view, the FISST definition~\eqref{eq:functional_derivative:Mahlers_definition} could be seen as a generalization of the Volterra derivative.
Although FISST definition \cite{Mahler-Statistic102:2013}, \cite[Appendix C.3]{Mahler-Book:2014} refers the reader to Volterra, note that multivariate limits can be split and interchanged under different conditions in the classic calculus used by Volterra and in the measure-theoretic approach of FISST. 

Different limiting procedures can be used that address the properties of Dirac delta more directly as follows.

\subsection{Definitions Specific to PGFLs via Distributions}\label{subsec:definition_Streit_distributions}
This construction follows~\cite[Appendix~B.4]{Streit-Book:2021} and~\cite{Streit_FunctionalDerivatives:2015}, that generalize the work from~\cite{Streit-PHD_ordinary_derivatives:2014}.
In fact, as indicated in~\cite{Streit-PHD_ordinary_derivatives:2014}, the construction is apparently inspired by use of Dirac deltas in the early treatment of point processes by Moyal~\cite{Moyal:1962}.
However, it should be emphasized that~\cite{Moyal:1962} used Dirac \emph{measures} to reconstruct a measure describing the point process (essentially a ``version'' of~\eqref{eq:probability_measure_RFS}), or the measure corresponding to the factorial moment~\eqref{eq:PGFL_to_MomentDensities}.
The treatment in~\cite{Moyal:1962} did not use density functions at all.
Note that the approaches presented in this and the following subsection originally were built upon the point process approach rather than FISST.

Assuming that $\X=\mathbb{R}^d$, the key concept employed in~\cite[Appendix~B.4]{Streit-Book:2021} and~\cite{Streit_FunctionalDerivatives:2015} is to use the theory of distributions.
Consider a family $\{ \DiracTestSeq{\lambda}{\mathbf{x}} \}_\lambda$ of functions parameterized by\footnote{
    To ensure that the unit of measurement of a density on $\X$ (now assumed to be $\mathbb{R}^{d}$) is $u^{-1}$, the unit of measurement of $\lambda$ is taken to be $u$.
} \mbox{$\lambda>0$}, converging to the Dirac delta centered at $\mathbf{x}\in\X$ as
\begin{align}
    \lim_{\lambda \searrow 0} \int_{\X} \DiracTestSeq{\lambda}{\mathbf{x}}(\mathbf{y}) \varphi(\mathbf{y}) \d \mathbf{y} = \varphi(\mathbf{x}) \, , \label{eq:family-of-functions:distributions}
\end{align}
for all \emph{test functions} $\varphi\in C_c^{\infty}(\X)$, where $C_c^{\infty}(\X)$ is the space of infinitely continuously-differentiable functions with a compact support on $\X$, see Supplement~\ref{appendix:Dirac_deltas}.\ref{appendix:Dirac_distribution}.
Such a family $\{\DiracTestSeq{\lambda}{\mathbf{x}}\}_{\lambda}$ then forms a \emph{test sequence}\footnote{
    ''Continuous`` families were used in~\cite{Streit_FunctionalDerivatives:2015}.
    Alternatively, discrete sequences $\{\DiracTestSeq{k}{\mathbf{x}}\}_{k\in\mathbb{N}}$ of functions and limit for $k\rightarrow+\infty$ were used in~\cite{Streit-Book:2021}.
} for the Dirac \emph{distribution} and will be used to provide a direction $g$.
A test function $\varphi$ will play the role of a Janossy density.

The definition of functional derivative of a PGFL $G_{\Xi}[h]$~\eqref{eq:def_PGFL} can be \emph{formally} given as the limit of directional derivatives 
\begin{subequations}\label{eq:Streit_def_distributions_all}
\begin{align}
    \frac{\delta G_{\Xi}}{\delta \mathbf{x}} [h]
    &= \lim_{\lambda \searrow 0} \frac{ \partial G_{\Xi} }{ \partial \DiracTestSeq{\lambda}{\mathbf{x}} } [h] \label{eq:Streit_def_distributions_a}\\
    &= \lim_{\lambda \searrow 0} \left[ \frac{\d}{\d \epsilon} G_{\Xi} [h +\epsilon \DiracTestSeq{\lambda}{\mathbf{x}} ] \right]_{\epsilon = 0}
    \, . \label{eq:Streit_def_distributions}
\end{align}
\end{subequations}
Functions $h$ and Janossy densities for which~\eqref{eq:Streit_def_distributions_all} is valid differ between~\cite[Appendix~B.4]{Streit-Book:2021} and~\cite{Streit_FunctionalDerivatives:2015}.
Nevertheless, the iterated version is defined by~\eqref{eq:def_functional_derivative_iterated_2}.
\\

\subsubsection{Definition according to~\cite[Appendix~B.4]{Streit-Book:2021}}\label{sec:definition-distributions-1}
In this definition, it is assumed that $h\in\Gamma(\X)$ and $\DiracTestSeq{\lambda}{\mathbf{x}} \in C_c^{\infty}(\X)$. 
Notice that for $h {\equiv} 1$, the term $(h + \epsilon \DiracTestSeq{\lambda}{\mathbf{x}})$ in~\eqref{eq:Streit_def_distributions} is not from $\Gamma(\X)$ and thus it is ill-defined for that case. 
This was addressed using the left-hand limit as~\cite[Appendix~B.4]{Streit-Book:2021} 
\begin{align}
    \frac{\delta G_{\Xi}}{\delta \mathbf{x}} [1] = \lim_{h\nearrow 1} \frac{\delta G_{\Xi}}{\delta \mathbf{x}} [h],
    \label{eq:Streit:derivative_at_1}
\end{align}
which ``defines'' the functional derivative at $h {\equiv} 1$.

\subsubsection*{Analysis of the definition}
alike for the preceding definitions~\ref{subsec:definition_ae} and~\ref{subsec:definition_set_derivatives}, the functional derivative under the current assumptions cannot be even Gâteaux as $\Gamma(\X)$ is not a linear space.
That is, the definition does not grant the chain rule.
Although for $\DiracTestSeq{\lambda}{\mathbf{x}} \in C_c^{\infty}(\X)$, the assumption $\X {=} \mathbb{R}^d$ is unnecessary, the space $\X$ cannot be an arbitrary \locallyCompactSecondCountableHausdorff{} space, see Supplement~\ref{appendix:Dirac_deltas}.\ref{appendix:Dirac_distribution}.
The same applies to the following definition.
Properties of the Janossy densities involved were neither analyzed explicitly in~\cite[Appendix~B.4]{Streit-Book:2021}, nor were properties of the iterated version.

\subsubsection{Definition according to~\cite[II.C]{Streit_FunctionalDerivatives:2015}}\label{sec:Degen-linear}
Let $\mathrm{H}(\X) \not\subset \Gamma(\X) $ be the set of real-valued bounded and Lebesgue-integrable functions.
Notice that $\mathrm{H}(\X)$ forms a linear space, and the constant function $h {\equiv} 1$ is not its element as it is not integrable over the entire $\X {=} \mathbb{R}^d$ unless $\referenceMeasure(\X)<+\infty$.
Consider the PGFL defined on $\mathrm{H}(\X)$, i.e., $G_{\Xi}:\mathrm{H}(\X)\rightarrow\mathbb{R}$ of the form~\eqref{eq:def_PGFL}.
Two cases regarding the family $\{ \DiracTestSeq{\lambda}{\mathbf{x}} \}_{\lambda}$ were discussed in~\cite{Streit_FunctionalDerivatives:2015}:
\def\caseGorS{$\mathrm{C}_1$}
\def\caseGen{$\mathrm{C}_2$}
\begin{itemize}
    \item[\caseGorS] $\{ \DiracTestSeq{\lambda}{\mathbf{x}} \}_{\lambda}$ is specified either via a Gaussian or indicator function, see Supplement~\ref{appendix:Dirac_deltas}.\ref{appendix:Dirac_distribution}.
    Janossy densities are then assumed to be from $\mathrm{B}(\X^n)$ for each $n {\in} \mathbb{N}$, where $\mathrm{B}(\X)$ is the set of continuous, bounded, and absolutely integrable functions on~$\X$.
    \item[\caseGen] $\{ \DiracTestSeq{\lambda}{\mathbf{x}} \}_{\lambda}$ is specified via any  absolutely integrable function, see Supplement~\ref{appendix:Dirac_deltas}.\ref{appendix:Dirac_distribution}.
    Janossy densities are then assumed to be from $C_c(\X^n)$ for each order $n {\in} \mathbb{N}$, where $C_c(\X^n)$ is the set of continuous functions with compact support in $\X$, see Supplement~\ref{appendix:topology}.
\end{itemize}

\subsubsection*{Analysis of the definition}
first, note that for a fixed value of $\lambda$ close to zero, it is likely that $\DiracTestSeq{\lambda}{\mathbf{x}} \notin \Gamma(\X)$, but there always exists sufficiently small $\epsilon {>} 0$ such that $(\epsilon \DiracTestSeq{\lambda}{\mathbf{x}}) \in \Gamma(\X)$. 
Next, notice that for a fixed $h$ such that $\sup_{\mathbf{x}\in\X} |h(\mathbf{x})|<1$ and a fixed $\DiracTestSeq{\lambda}{\mathbf{x}}$, the term $G_{\Xi}[h {+} \epsilon\DiracTestSeq{\lambda}{\mathbf{x}}]$ as a function of $\epsilon$ is analytic in some open region of the complex plane that contains the origin~\cite[Sec.~4]{Moyal:1962}.
The ordinary derivative in~\eqref{eq:Streit_def_distributions} thus exists, but both $h$ and the direction $\DiracTestSeq{\lambda}{\mathbf{x}}$ must be fixed.
The properties of the functional derivative~\eqref{eq:Streit_def_distributions} w.r.t.~$h$ and $\DiracTestSeq{\lambda}{\mathbf{x}}$ were not studied further\footnote{
    According to \cite{Streit_FunctionalDerivatives:2015}, the fact that $G_{\Xi}[h+\epsilon\DiracTestSeq{\lambda}{\mathbf{x}}]$ is analytic as a function of $\epsilon$ around the origin of $\mathbb{C}$ implyes that the resulting functional derivative is Fréchet.
    This statement is valid only for fixed $h$ and $\DiracTestSeq{\lambda}{\mathbf{x}}$, i.e., the \emph{ordinary} derivative of $G_{\Xi}[h+\epsilon\DiracTestSeq{\lambda}{\mathbf{x}}]$ w.r.t.~$\epsilon$ is Fréchet.
    This statement is, however, missing the point as $\mathbb{C}$ is a simple example of a Banach space and analytic functions are Fréchet differentiable trivially.
} in~\cite{Streit_FunctionalDerivatives:2015}.
As a result, it is unclear whether~\eqref{eq:Streit_def_distributions_all} under the current assumptions grants the use of the chain rule.
If so, note that~\eqref{eq:Streit_def_distributions_all} must be a Chain derivative instead of Fréchet since the space of distributions is not a normed space as it is a topological vector space whose topology is not even metrizable~\cite[p.~156]{Rudin:FunctionalAnalysis:1973}. 
\\
Note that for~\eqref{eq:Streit_def_distributions_all} to be well-defined within the theory of distributions, it cannot depend on a particular choice of the family $\{\DiracTestSeq{\lambda}{\mathbf{x}}\}_{\lambda}$.
When any particular $\{\DiracTestSeq{\lambda}{\mathbf{x}}\}_{\lambda}$ specification is fixed, the Dirac delta is approximated in a certain way that may lead to inconsistent results (whithin the calculus of distributions) when considered valid on the ``larger'' set $B(\X) {\supset} C_c(\X)$, i.e., for Janossy densities being from $B(\X^n)$, $n {\in} \mathbb{N}$.
As a result, \eqref{eq:Streit_def_distributions_all} is ill-defined under the assumption \caseGorS{} as the modeled Dirac delta lies off the standard theory of distributions~\cite[Chapter~9]{Folland:1999}.
On the other hand, for \caseGen{}, the Janossy densities are posed to have a compact support, and so they could not even be composed of Gaussians.
\\
With both \caseGorS{} and \caseGen{}, this definition is applicable and the resulting functional derivative is defined point-wise, but it is not defined for $h {\equiv} 1$, $\X$ cannot be arbitrary \locallyCompactSecondCountableHausdorff{} space and it is unclear whether the chain rule can be used.
Properties of the iterated version were not studied in~\cite{Streit_FunctionalDerivatives:2015}.
\\[-0.2cm]

A connection between the definition~\eqref{eq:Streit_def_distributions_all} and the FISST definition~\eqref{eq:functional_derivative:Mahlers_definition} can be established by specifying the test sequence using the indicator function~\cite{Streit_FunctionalDerivatives:2015}.

\subsection{Definition Using Secular Functions}\label{subsec:definition_Streit_Secular}
The main idea of the Secular method of~\cite{Streit-PHD_ordinary_derivatives:2014,Streit_FunctionalDerivatives:2015}\cite[Appendix~B.5]{Streit-Book:2021} is to interchange the limits and the derivative in~\eqref{eq:Streit_def_distributions}, formally leading to
\begin{align}
    \!\!
    \frac{\delta G_{\Xi}}{\delta \mathbf{x}} [h]
    =& \bigg[ \frac{\d}{\d \epsilon} \, \underbrace{ \lim_{\lambda \searrow 0}  G_{\Xi} [h +\epsilon \DiracTestSeq{\lambda}{\mathbf{x}} ] }_{S_{\Xi,h}(\epsilon)} \, \bigg]_{\epsilon = 0}
    = \frac{\d S_{\Xi,h}}{\d \epsilon} (0),
    \label{eq:secular-method}
\end{align}
where the function $S_{\Xi,h}(\epsilon)$ is called the secular function.
Notice that once the secular function is known, ordinary derivatives suffice to establish the functional derivative.
This definition can be considered under various assumptions detailed in the preceding definitions.

\subsubsection*{Analysis of the definition}
first, consider the functional $G_{\Xi}$ is defined on $\Gamma(\X)$.
As opposed to the previous definition, the behavior of $G_{\Xi}[h {+} \epsilon \DiracTestSeq{\lambda}{\mathbf{x}}]$ is to be studied for variable $\lambda$, while $\epsilon$ is kept constant.
Recall that there always exists sufficiently small $\epsilon {>} 0$ such that $(\epsilon \DiracTestSeq{\lambda}{\mathbf{x}}) \in \Gamma(\X)$, as far as $\lambda {>} 0$ is fixed.
Fixing $\epsilon {>} 0$ to be arbitrarily small, however, leaves enough room for $\lambda {>} 0$ small enough to yield a function $\epsilon \DiracTestSeq{\lambda}{\mathbf{x}}$ that is not from $\Gamma(\X)$.
As a result, secular function $S_{\Xi,h}$ is ill-defined unless $\epsilon {=} 0$ and its ordinary derivative thus does not exist (even at $\epsilon {=} 0$).

As discussed in~\cite{Streit_FunctionalDerivatives:2015}, considering $G_{\Xi}$ defined on $H(\X)$, the above problems vanish.
On the other hand, essentially the same arguments as described in Section~\ref{sec:Degen-linear} are still valid.\\[-0.2cm]

To the best of our knowledge, all the definitions of functional derivatives in the context of MOT were covered.
In summary, neither of the presented definitions addresses all the requirements formulated in Section~\ref{sec:problem-statement}.
Most importantly, no definition was ever shown to \emph{grant} the use of the chain rule.
The definitions originating within the FISST background~\ref{subsec:definition_ae} and \ref{subsec:definition_set_derivatives} never detailed assumptions on the Janossy densities for which the definitions are valid.
In that regard, the subsequent definitions~\ref{subsec:definition_Streit_distributions} and~\ref{subsec:definition_Streit_Secular} made a step forward.
In the following, we further generalize this result and present a novel and rigorous definition of the functional derivative.
It could be viewed as a generalization of the definition~\ref{sec:Degen-linear} with \caseGorS{}, but using the measure theory instead of the theory of distributions.

Note that~\cite[p.~11]{Mahler-Statistic103:2019} pointed out that the FISST (Volterra) definition ``{permits the direct derivation of density functions without resorting to measures}''.
To the best of our knowledge, however, such derivation is unavailable in the literature so far.
The following section is the first step in this direction.

\section{Proposed Solution}\label{sec:proposed}
To overcome issues of the existing definitions described above, this section proposes a redefinition of the PGFL.
The redefined functional has measures as its input instead of functions.
This is motivated by the fact that the Dirac delta can be conveniently treated as a measure, see Supplement~\ref{appendix:Dirac_deltas}.
The mathematical machinery of measures on locally compact and Hausdorff (\locallyCompactHausdorff{}) spaces reviewed in Supplement~\ref{appendix:Radon_Measures} then helps to overcome some of the aforementioned issues.  

Note that the set integral in~\eqref{eq:def_set_Integral} is defined using the reference Radon measure $\referenceMeasure$ introduced in Section~\ref{sec:Background:TheBaseSpace}.
In the following, the notation~\eqref{eq:def_set_Integral} is generalized so that any measure $\eta$ on $\X$ can be used, as
\begin{align}
    &\hspace{3.5cm} \int_{\X} f(X) \eta^{\delta X} \hspace{0.2cm} \triangleq \notag\\[-0.5cm]
    &\hspace{0.0cm} \sum_{n=0}^{+\infty} \frac{1}{n!} \overbrace{ \int_{\X}\cdots\int_{\X} }^{n \text {-times}} f(\{ \mathbf{x}_1,\dots,\mathbf{x}_n \}) \eta(\d \mathbf{x}_1) \cdots \eta(\d \mathbf{x}_n) \, . 
\end{align}
With this notation, the original set integral~\eqref{eq:def_set_Integral} is recovered as $\int_{\X} f(X) \delta X = \int_{\X} f(X) \referenceMeasure^{\delta X}$, using the right hand side of~\eqref{eq:integral-wrt-reference-measure-notation}.

\subsection{Probability Generating Functional with Measure Inputs}\label{sec:proposed-PGFM-def}
In this section, the PGFL $G_{\Xi}[h]$~\eqref{eq:def_PGFL} is redefined such that its domain is the \emph{linear normed space} of \emph{complex Radon} measures $M(\X)$, which is the dual space of \emph{continuous complex-valued functions vanishing at infinity} denoted with $\mathrm{C}_{0}(\X)$ as discussed in Supplement~\ref{appendix:Radon_Measures}.
To distinguish the original PGFL $G_{\Xi}[h]$~\eqref{eq:def_PGFL} from the proposed one, the latter is further called the \emph{probability generating functional with measure inputs} (PGFM) $G_{\Xi}:M(\X)\rightarrow \mathbb{C}$ and is defined as
\begin{align}
    G_{\Xi}[\eta] \triangleq \int_{\X} p_{\Xi}(X) \eta^{\delta X} \, , \label{eq:def:PGFM} 
\end{align}
where $p_{\Xi}(\{ \mathbf{x}_1,\dots,\mathbf{x}_n \}) {\in} C_0(\X^n)$ is a fixed Janossy (real-valued non-negative symmetric) density for each order $n {\in} \mathbb{N}$.
Note that functions from $C_0(\X^n)$ are defined point-wise everywhere, and so are the Janossy densities.
Also note that such functions are automatically bounded~\cite[p.~132]{Folland:1999}.

For $\eta$ being the reference measure $\referenceMeasure$, the PGFM~\eqref{eq:def:PGFM} becomes $G_{\Xi}[\referenceMeasure] {=} 1$, i.e., setting $\eta {=} \referenceMeasure$ behaves as if $h {\equiv} 1$ for the PGFL definition.
However, it should be emphasized that the reference Radon measure $\referenceMeasure$ does not necessarily belong to the \emph{linear} space $M(\X)$.
In fact, even the Lebesgue measure on $\X {=} \mathbb{R}^d$ does not belong to $M(\mathbb{R}^d)$, since for the Lebesgue measure we have $\referenceMeasure(\mathbb{R}^d) {=} {+}\infty$.
For the Dirac measure $\delta_{\mathbf{x}}$ centered at $\mathbf{x} \in \X$, however, we have the following.\\[-0.2cm]

\begin{proposition}\label{proposition:dirac-measure-is-Radon}
The Dirac measure is a Radon measure assuming $\X$ is locally compact Hausdorff (\locallyCompactHausdorff{}) space, i.e., $\delta_{\mathbf{x}} \in M(\X)$.
The proof is given in Appendix~\ref{appendix:Dirac_deltas-proof}.
\hfill$\square$\\[-0.2cm]
\end{proposition}

Note that the second countability is not needed in the proof, and the statement is valid for $\X$ being \locallyCompactSecondCountableHausdorff{} as well.
With this result, the Dirac delta interpreted as a measure can be legitimately substituted into the PFGM~\eqref{eq:def:PGFM} and thus yields a directional derivative in the \emph{standard} sense as opposed to the definitions reviewed in Section~\ref{sec:definitions}.

\subsection{The Proposed Functional Derivative of PGFM}\label{subsec:proposed-definition-itself}
The first-order \emph{directional} derivative of the PGFM $G_{\Xi}[\eta]$~\eqref{eq:def:PGFM} in the direction $\nu\in M(\X)$ is defined as
\begin{align}
    \frac{\partial G_{\Xi}}{ \partial \nu } [\eta] &\triangleq \lim_{\epsilon \rightarrow 0} \frac{ G_{\Xi}[\eta + \epsilon\nu] - G_{\Xi}[\eta] }{ \epsilon } \, .
    \label{eq:PGFM:directional:derivative}
\end{align}

The first-order \emph{functional} derivative of the PGFM is defined as the directional derivative in the direction of the Dirac delta measure.
That is, the definitions~\eqref{eq:def_functional_derivative} and~\eqref{eq:PGFM:directional:derivative} become
\begin{align}
    \frac{\delta G_{\Xi}}{ \delta\mathbf{x} } [\eta] \triangleq \frac{\partial G_{\Xi}}{ \partial \delta_{\mathbf{x}} } [\eta].
    \label{eq:PGFM:functional:derivative:firt_order}
\end{align}
Notice that $(\eta+\epsilon \delta_{\mathbf{x}})\in M(\X)$, since $M(\X)$ is a \emph{linear} space.
The higher-order functional derivatives are defined iteratively using~\eqref{eq:def_functional_derivative_iterated}, where the function $h$ is formally substituted with the measure $\eta {\in} M(\X)$ and $\delta_{\mathbf{x}_n} {\in} M(\X)$ is the Dirac measure centered at $\mathbf{x}_n {\in} \X$, i.e.,
\begin{align}
    \!\! & \tfrac{ \delta G_{\Xi} }{ \delta \{ \mathbf{x}_1, \dots, \mathbf{x}_m\} }[\eta] \! \triangleq \! \tfrac{\partial^m G_{\Xi}}{ \partial \delta_{\mathbf{x}_1}\cdots \partial \delta_{\mathbf{x}_m} } [\eta] \, , \label{eq:PGFM:functional:derivative:mth_order}
    \\
    \!\! & \tfrac{\partial^m G_{\Xi}}{ \partial \nu_1\cdots \partial \nu_m } [\eta] \! \triangleq \! \lim_{\epsilon \rightarrow 0} \! \frac{ \tfrac{ \partial^{m{-}1} G_{\Xi} }{ \partial \nu_1 \cdots \partial \nu_{m{-}1} } [\eta {+} \epsilon\nu_m] {-} \tfrac{ \partial^{m{-}1} G_{\Xi} }{ \partial \nu_1 \cdots \partial \nu_{m{-}1} }[\eta] }{ \epsilon }. \!\! \label{eq:PGFM:mth-order-directional:derivative}
\end{align}

\subsection{Analysis of the Proposed Definition}
The key properties of the proposed definition of functional derivative are sumarized in \textit{Corollaries}~\ref{corollary:analysis-of-first-order-func-der} and~\ref{corollary:analysis-of-mth-order-func-der}, that are granted by the following Proposition.\\[-0.2cm]

\begin{proposition}\label{proposition:first-order-Fréchet}
Let $\X$ be \locallyCompactHausdorff{} topological space, measures $\eta,\nu_1,\dots,\nu_m {\in} M(\X)$, number $p_0 {\in} \mathbb{C}$ and sequence of functions\footnote{
    The functions $p_n(\mathbf{x}_1,\dots,\mathbf{x}_n)$ need not be symmetric for the proof.
} $p_n(\mathbf{x}_1,\dots,\mathbf{x}_n) \in C_0(\X^n)$ $\forall n {\in} \mathbb{N}$ such that a finite uniform bound $K$ for all finite cardinalities exists, i.e.,
\begin{align}
    K \triangleq \sup_{ \mathbf{x}_1,\dots,\mathbf{x}_n \in C_0(\X^n), \, n\in\mathbb{N}_0 } \ p_n( \mathbf{x}_1,\dots,\mathbf{x}_n ) \ < \ +\infty.
    \label{app:eq:bound-for-all-cardinalities}
\end{align}
Let $G {:} M(\X) {\rightarrow} \mathbb{C}$ be functional defined as
\begin{align}
    G[\eta] = \sum_{n=0}^{+\infty} \frac{1}{n!} \underbrace{ \int_{\X}\!\cdots\!\int_{\X} }_{n \text {-times}} p_n( \mathbf{x}_1,\dots,\mathbf{x}_n ) \eta(\d \mathbf{x}_1) \cdots \eta(\d \mathbf{x}_n) \, .
    \notag \\[-0.6cm]
    \label{eq:generic-functional-on-M}
\end{align}
Then the $m$-th order directional derivative~\eqref{eq:PGFM:mth-order-directional:derivative}, $m {\in} \mathbb{N}$, of $G[\eta]$~\eqref{eq:generic-functional-on-M} in the directions $\nu_1,\dots,\nu_m$ is Fréchet.
The proof is given in Appendix~\ref{appendix:first-order-Frechet-proof}.
\hfill$\square$\\[-0.2cm]
\end{proposition}

The forms of the first and general $m$-th order directional derivatives of $G[\eta]$~\eqref{eq:generic-functional-on-M} are given in~\eqref{app:eq:PGFM:directional-derivative:general} and~\eqref{app:eq:PGFM:directional-derivative:general:mth-order}, respectively.
It follows directly that the directional derivatives~\eqref{eq:PGFM:directional:derivative}, \eqref{eq:PGFM:mth-order-directional:derivative} of a PGFM $G_{\Xi}[\eta]$~\eqref{eq:def:PGFM} are Fréchet and thus obey the chain and product differentiation rules~\cite[pp.~312--313]{LusternikSobolev:ElementsOfFunctionalAnalysis:1975}.
Further implications for the special case of functional derivatives~\eqref{eq:PGFM:functional:derivative:firt_order}, \eqref{eq:PGFM:functional:derivative:mth_order} are discussed in the following Corollaries.\\[-0.2cm]

\begin{corollary}\label{corollary:analysis-of-first-order-func-der}
The first-order functional derivative~\eqref{eq:PGFM:functional:derivative:firt_order} of a PGFM $G_{\Xi}[\eta]$~\eqref{eq:def:PGFM} is Fréchet, has the form
\begin{align}
    & \frac{\delta G_{\Xi}}{ \delta\mathbf{x} } [\eta] = 
    \sum_{n=1}^{+\infty} \frac{1}{n!} \sum_{i=1}^n \overbrace{ \int_\X \dots \int_\X }^{ n\text{-times} } p_{\Xi}(\{\mathbf{x}_1,\dots,\mathbf{x}_n\}) \notag\\
    & \hspace{0.5cm} \cdot \eta(\d\mathbf{x}_1) \cdots \eta(\d\mathbf{x}_{i-1}) \delta_{\mathbf{x}}(\d\mathbf{x}_i)\eta(\d\mathbf{x}_{i+1}) \cdots \eta(\d\mathbf{x}_n) \, ,
    \label{app:eq:PGFM:functional-derivative}
\end{align}
and the following holds.
\begin{itemize}
    \item For $\eta\equiv 0$ being the zero measure, Eq.~\eqref{app:eq:PGFM:functional-derivative} becomes
    \begin{align}
        \frac{\delta G_{\Xi}}{ \delta\mathbf{x} } [0] = p_{\Xi}(\{\mathbf{x}\}) \, ,
        \label{eq:proposed-at-zero}
    \end{align}
    and thus PGFM recovers the density like PGFL does recover~\eqref{eq:PGFL_to_PDF} for $X=\{\mathbf{x}\}$.
    \item When~\eqref{app:eq:PGFM:functional-derivative} is restricted to \emph{positive} measures, i.e., neither signed nor complex measures, the integration admits the Fubini--Tonelli Theorem~\cite[Thm.~2.37]{Folland:1999}, see Appendix~\ref{appendix:nth-order-Frechet-proof}. 
    Integrals in~\eqref{app:eq:PGFM:functional-derivative} can then be reordered, and
    \begin{align}
        \frac{\delta G_{\Xi}}{ \delta\mathbf{x} } [\eta] = \int_\X p_{\Xi}(X \cup \{\mathbf{x}\}) \eta^{\delta X} \, .
        \label{eq:proposed-derivative-restricted:set-integral-notation}
    \end{align}
    Note that positive measures $\eta$ no longer form a linear space\footnote{
        For instance, $(-1)\cdot \eta$ is not a \emph{positive}, but rather a \emph{signed} measure.
    } that is required for the Fréchet (and also chain differentiability) and thus for the chain rule to hold.
    \item If the reference measure $\referenceMeasure$ is \emph{positive} and belongs to $M(\X)$, Eq.~\eqref{app:eq:PGFM:functional-derivative} can be evaluated for $\eta {=} \referenceMeasure$ yielding
    \begin{align}
        \frac{\delta G_{\Xi}}{ \delta\mathbf{x} } [\lambda] =
        D_{\Xi}(\{\mathbf{x}\}) \, ,
        \label{eq:proposed-at-lambda}
    \end{align}
    and thus PGFM recovers factorial moments as the PGFL does recover~\eqref{eq:PGFL_to_MomentDensities} for $X=\{\mathbf{x}\}$.
    If $\referenceMeasure$ is chosen as a generalization of the Lebesgue measure for the \locallyCompactSecondCountableHausdorff{} space $\X$, it must hold that $\referenceMeasure(\X)<+\infty$.
    Using the Lebesgue measure on $\mathbb{R}^d$, the space $\X$ can be any bounded measurable subset of $\mathbb{R}^d$.
    If Janossy densities are assumed Gaussian, they must be truncated.
    \hfill$\square$\\[-0.2cm]
\end{itemize}
\end{corollary}

\begin{corollary}\label{corollary:analysis-of-mth-order-func-der}
The $m$-th order functional derivative~\eqref{eq:PGFM:functional:derivative:mth_order} of a PGFM $G_{\Xi}[\eta]$~\eqref{eq:def:PGFM} with respect to $\{\mathbf{y}_1,\dots,\mathbf{y}_m\}\subset \X$ is Fréchet, has the form
\begin{align}
    & \frac{\delta G_{\Xi}}{ \delta \{ \mathbf{y}_1, \dots, \mathbf{y}_m \} } [\eta] = 
    \sum_{n=m}^{+\infty} \frac{1}{n!}
        \ \sum_{ 1 \leq i_1 < \dots < i_m \leq n } \
        \sum_{ \pi \in \mathrm{Sym}(m) } \notag \\[-0.1cm]
    & \overbrace{ \textstyle \int_\X \dots \int_\X }^{ n\text{-times} } p_{\Xi}(\{ \mathbf{x}_1,\dots,\mathbf{x}_n \}) 
        \cdot \eta(\d\mathbf{x}_1) \cdots \notag \\
    &  \hspace{0.6cm} \cdots \eta(\d\mathbf{x}_{i_1-1}) \delta_{ \mathbf{y}_{\pi(1)} }(\d\mathbf{x}_{i_1}) \eta(\d\mathbf{x}_{i_1+1}) \cdots \notag \\
    & \hspace{0.4cm} \cdots \eta(\d\mathbf{x}_{i_m-1}) \delta_{ \mathbf{y}_{\pi(m)} }(\d\mathbf{x}_{i_m}) \eta(\d\mathbf{x}_{i_m+1}) 
        \cdots \eta(\d\mathbf{x}_n) \, ,
    \label{app:eq:PGFM:functional-derivative:mth-order}
\end{align}
and the following holds.
\begin{itemize}
    \item For $\eta\equiv 0$ being the zero measure, Eq.~\eqref{app:eq:PGFM:functional-derivative:mth-order} becomes
    \begin{align}
        \frac{\delta G_{\Xi}}{ \delta \{ \mathbf{y}_1, \dots, \mathbf{y}_m \} } [0] = p_{\Xi}(\{\mathbf{y}_1,\dots,\mathbf{y}_m\}) \, ,
        \label{eq:proposed-at-zero:mth-order}
    \end{align}
    and thus satisfies~\eqref{eq:PGFL_to_PDF} for $X=\{\mathbf{y}_1,\dots,\mathbf{y}_m\}$.
    \item When~\eqref{app:eq:PGFM:functional-derivative:mth-order} is restricted \emph{positive} measures, the Fubini\discretionary{--}{-}{--}Tonelli Theorem applies~\cite[Theorem~2.37]{Folland:1999}.
    As a result, 
    \begin{align}
        \! \frac{\delta G_{\Xi}}{ \delta Y } [\eta] = \int_\X p_{\Xi}(X \cup Y) \eta^{\delta X} \, .
        \label{eq:proposed-derivative-restricted:set-integral-notation:mth-order}
    \end{align}
    \item If the reference measure $\referenceMeasure$ is \emph{positive} and belongs to $M(\X)$, Eq.~\eqref{app:eq:PGFM:functional-derivative:mth-order} can be evaluated for $\eta {=} \referenceMeasure$ yielding
    \begin{align}
        \frac{\delta G_{\Xi}}{ \delta Y } [\lambda] =
        D_{\Xi}( Y ) \, ,
        \label{eq:proposed-at-lambda:mth-order}
    \end{align}
    and thus satisfies~\eqref{eq:PGFL_to_MomentDensities}. 
    \hfill$\square$\\[-0.2cm]
\end{itemize}
\end{corollary}

Apparently, both~\eqref{app:eq:PGFM:functional-derivative}, \eqref{app:eq:PGFM:functional-derivative:mth-order} exist for all $\eta {\in} M(\X)$, any set $\{\mathbf{y}_1,\dots,\mathbf{y}_m\} {\subset} \X$ of $m$ elements ($m{=}1$ in case of \eqref{app:eq:PGFM:functional-derivative}), and any symmetric density $p_{\Xi}(X) {\in} C_0(\X^{|X|})$ with $p_{\Xi}(\emptyset) {\in} [0,1]$.
Arguably, both~\eqref{app:eq:PGFM:functional-derivative} and~\eqref{app:eq:PGFM:functional-derivative:mth-order} appear as involved mathematical expressions that result in the expressions known from the literature \eqref{eq:proposed-derivative-restricted:set-integral-notation}, \eqref{eq:proposed-at-lambda} and~\eqref{eq:proposed-derivative-restricted:set-integral-notation:mth-order}, \eqref{eq:proposed-at-lambda:mth-order} only when restricted to a specialized set of measures.
The user should keep in mind that~\eqref{app:eq:PGFM:functional-derivative} and~\eqref{app:eq:PGFM:functional-derivative:mth-order} cannot be used directly as a definition in practice for reasons discussed in Section~\ref{subsec:definition_for_PGFLS}.

Remind that the proposed derivative applies to both random finite sets and point processes formalisms appearing in MOT literature (see Supplement~\ref{appendix:RFS_notes}), and that only widely accepted mathematical techniques were used in the construction.
To avoid any mathematically inconsistent or complicated treatment, the development of the functional derivative in this paper stops here.
Several aspects are left to be addressed in the future and are discussed below.

\newcommand{\Prb}[1]{OP$_#1$}
\subsection{Open Problems}
As revealed above, the proposed definition of functional derivatives yields some of the key properties that are missing in the alternative definitions in the literature.
Namely, the derivative was proven to be Fréchet and ranges of validity were discussed.
To demonstrate that MOT algorithms developed using functional derivative techniques are mathematically valid in the proposed sense, however, the performance of the following exhaustive procedure is needed:
\emph{(i)}~consider the PGFL $G_{\Xi}[h]$~\eqref{eq:def_PGFL} at hand as the PGFM $G_{\Xi}[\eta]$~\eqref{eq:def:PGFM}, \emph{(ii)}~compute its directional derivatives~\eqref{eq:PGFM:directional:derivative} and \eqref{eq:PGFM:mth-order-directional:derivative}, \emph{(iii)}~evaluate the resulting expressions at the directions $\delta_{\mathbf{x}_1}, \dots, \delta_{\mathbf{x}_m}$ and $\eta$ being a positive measure, and finally \emph{(iv)}~check that the resulting expressions match with the conclusions made during the MOT algorithm development before.
If \emph{(iv)} fails it may not necessarily be incorrect. 
This motivates the following open problems \Prb{1}--\Prb{6}.

\begin{itemize}
    \item[\Prb{1}]
    Although both~\eqref{eq:PGFM:directional:derivative} and \eqref{eq:PGFM:mth-order-directional:derivative} obey chain and product differentiation rules, the formally-determined rules~\cite[Section~3.5]{Mahler-Book:2014} apply only after restricting the resulting expressions (generally of the form~\eqref{app:eq:PGFM:directional-derivative:general},\eqref{app:eq:PGFM:directional-derivative:general:mth-order}) to specialized directions $\delta_{\mathbf{x}_1},\dots,\delta_{\mathbf{x}_m}$ and $\eta$ to specialized set of measures.
    That is, the rules from~\cite[Section~3.5]{Mahler-Book:2014} do not apply directly to the proposed derivative and working with~\eqref{eq:PGFM:directional:derivative}--\eqref{eq:PGFM:functional:derivative:mth_order} may be challenging, although provably rigorous unlike with the other definitions.
    \item[\Prb{2}]
    Eq.~\eqref{eq:PGFL_to_MomentDensities}, i.e., Eq.~\eqref{eq:proposed-at-lambda:mth-order}, \eqref{eq:proposed-at-lambda},  are necessary for practice. However, for a general \locallyCompactSecondCountableHausdorff{} space $\X$  Eq.~\eqref{eq:proposed-at-lambda:mth-order}, \eqref{eq:proposed-at-lambda} may not hold as $\referenceMeasure$ is not necessarily from $M(\X)$.
    \item[\Prb{3}]
    The proposed directional and functional derivatives depend on the order of the directions in which the PGFM is being differentiated. That is, to avoid misinterpretation, the notation ``$\{\,\}$'' in $\tfrac{ \delta }{ \delta \{ \mathbf{x}_1, \dots, \mathbf{x}_m \} }$ in~\eqref{eq:PGFM:functional:derivative:mth_order} and~\eqref{app:eq:PGFM:functional-derivative:mth-order} shall be taken with care as was done in~\eqref{eq:proposed-at-zero:mth-order}, \eqref{eq:proposed-derivative-restricted:set-integral-notation:mth-order} and \eqref{eq:proposed-at-lambda:mth-order}.
    A generalization of the Fubini\discretionary{--}{-}{--}Tonelli Theorem for complex measures\footnote{
        Note that the symmetry of $p_{\Xi}(X)$ plays no role in this issue. 
        For general measures $\nu_1,\nu_2\in M(\X)$, it is possible that
        \begin{align}
        \textstyle 
            \int_\X \int_\X \nu_1(\d \mathbf{x}_1)\nu_2(\d \mathbf{x}_2)
            \neq 
            \int_\X \int_\X \nu_2(\d \mathbf{x}_1)\nu_1(\d \mathbf{x}_2) \,.
        \end{align}
    }, if possible, would help establishing order-independency in~\eqref{eq:PGFM:functional:derivative:mth_order}, \eqref{app:eq:PGFM:functional-derivative:mth-order}.
    \item[\Prb{4}]
    Although $p_{\Xi}(X) {\in} C_0(\X^{|X|})$ can be composed, e.g., of Gaussians, they cannot be composed of continuous uniform densities as these are not elements of $C_0(\X^{|X|})$.
    \item[\Prb{5}]
    The MOT motion and measurement models themselves may involve Dirac deltas, such as in the trajectory formulation~\cite{SetsOfTrajectories:2016,TPMBM-Continuity:2018,PHD-Trajectory:2019}.
    In that case, the corresponding densities are not contained in $C_0(\X^{|X|})$.
    \item[\Prb{6}]
    Functionals of several variables are left unaddressed.
\end{itemize}
Further work is required to analyze \emph{if} and \emph{how} the above issues can be addressed, thus making functional derivatives more handy besides being mathematically rigorous.
It should be noted that the proposed derivative satisfies most of the desired properties discussed in Section~\ref{sec:problem-statement} and thus \emph{outperforms} the existing definitions reviewed in Section~\ref{sec:definitions}.
Comparison of the properties granted by the various definitions is given in Table~\ref{tab:properties}.

\newcommand{\Satisfies}{\ding{51}}
\newcommand{\NotSatisfies}{\ding{55}}
\newcommand{\Unknown}{unclear}
\newcommand{\notApplicable}{N/A}
\newcommand{\mayHold}{may hold in some cases}
\begin{table*}[t]
    \centering
    \caption{
        Comparison of various functional derivative definitions.
    }
    \label{tab:properties}
    \begin{tabular}{lcccccc}
    \toprule
    definition
     & admissible $\X$
     & admissible $p_{\Xi}(X)$
     & factorial moments recovery
     & chain rule 
     & user friendliness \\ \midrule
     heuristic \eqref{eq:def_functional_derivative}-\eqref{eq:def_functional_derivative_iterated} & any \locallyCompactSecondCountableHausdorff{} & not specified & \mayHold & \NotSatisfies{} & most friendly \\
     bypassing limit procedure \ref{subsec:definition_for_PGFLS} & any \locallyCompactSecondCountableHausdorff{} & unconstrained & \Satisfies{} & \NotSatisfies{} & extremely unfriendly \\
     almost everywhere \ref{subsec:definition_ae} & any \locallyCompactSecondCountableHausdorff{} & abs. continuous & \Satisfies{} & \NotSatisfies{} & very unfriendly \\
     FISST via set derivative \ref{subsec:definition_set_derivatives} & any \locallyCompactSecondCountableHausdorff{} & abs. continuous, point-wise & \Satisfies{} & \NotSatisfies{} & somewhat friendly \\
     using distribution theory \ref{sec:definition-distributions-1} & $\mathbb{R}^d$ & not specified & \mayHold & \NotSatisfies{} & somewhat unfriendly \\
     using distribution theory \ref{sec:Degen-linear}\caseGorS{} & $\mathbb{R}^d$ & $\mathrm{B}(\X^{|X|})$ thus point-wise & \mayHold & \Unknown{} & somewhat unfriendly \\
     using distribution theory \ref{sec:Degen-linear}\caseGen{} & $\mathbb{R}^d$ & $\mathrm{C}_c(\X^{|X|})$ thus point-wise & \mayHold & \Unknown{} & somewhat unfriendly \\
     using secular method \ref{subsec:definition_Streit_Secular} & $\mathbb{R}^d$ & not specified & \NotSatisfies{} & \NotSatisfies{} & very friendly \\
     \rowcolor{gray!20}
     proposed \ref{subsec:proposed-definition-itself} & any \locallyCompactSecondCountableHausdorff{} & $C_0(\X^{|X|})$ thus point-wise & on positive measures & \Satisfies{} & friendly, challenging \\
     \bottomrule
    \end{tabular}
\end{table*}

\section{Conclusion}\label{sec:conclusion}
This paper focused on theoretical aspects of multi-object tracking, namely, on the lack of mathematical rigor regarding the use of functional derivatives.
Currently existing definitions of functional derivatives were shown to fail to fulfill several requirements.
Some are impractical, others have unclear or limited range of validity, or they are defined for too restrictive base (state or measurement) spaces only.
It is important to note that current definitions do not adequately provide for the chain differentiation rule. 

As a first step to address this inconvenience, this paper proposed a definition based on measure theory on locally compact and Hausdorff spaces.
Namely, the Dirac delta used in the definition is mathematically constructed as a measure, i.e., the Dirac measure.
The functional derivative of any finite order, as defined, is Fréchet, allowing the application of the chain rule and other differentiation rules.
While the proposed definition is mathematically rigorous, the corresponding analytical expressions may be too complex for practical use, which necessitates further investigation.
The list of open problems leaves room for future work.

\appendices
\renewcommand\thesubsection{\thesection\arabic{subsection}}
\renewcommand\thesubsectiondis{\thesection.\arabic{subsection}}

\section{Proof that the Dirac Measure is a Radon Measure (Proposition~\ref{proposition:dirac-measure-is-Radon})}\label{appendix:Dirac_deltas-proof}

Assuming $\X$ is \locallyCompactHausdorff{} and $\Sigma$ is the corresponding Borel \mbox{$\sigma$-algebra}, Proposition~\ref{proposition:dirac-measure-is-Radon} from Section~\ref{sec:proposed-PGFM-def} claims that that $\delta_\mathbf{x} \in M(\X)$.
Apparently, $\delta_\mathbf{x}$ is a (complex) Borel measure for any $\mathbf{x}\in \X$, and its norm is finite $\|\delta_\mathbf{x}\| = 1$. 
Proposition~\ref{proposition:dirac-measure-is-Radon} then follows from the fact that every complex Borel measure on a \locallyCompactHausdorff{} space is Radon~\cite[p.~222]{Folland:1999}.
Nevertheless, below, we give an alternative proof following the very definition of Radon measures along the lines of~\cite{Folland:1999}.

\textit{Proof of Proposition~\ref{proposition:dirac-measure-is-Radon} from Section~\ref{sec:proposed-PGFM-def}}\label{app:subsubsec:proof-of-proposition-Dirac}
First, observe that $\delta_\mathbf{x}$ is finite and real-valued and thus also a complex Borel measure.
It is easy to see that $\delta_\mathbf{x}^{+} = \delta_\mathbf{x}$ and $\delta_\mathbf{x}^{-}=0$.

Now, we will verify that $\delta_\mathbf{x}$ is outer regular on Borel sets. 
Let $E\in \mathfrak{B}(\X, \mathcal{T})$.
If $\mathbf{x}\in E$, then 
\begin{align}
    \delta_\mathbf{x}(E) = 1 = \inf 1 = \inf\{\delta_\mathbf{x}(U) \! : U {\supset} E {\ni} \mathbf{x}, U \text { open }\} \, .
\end{align}
On the other hand, if $x\not\in E$, then 
\begin{subequations}
\begin{align}
    0=\delta_\mathbf{x}(E) \leq \inf\{\delta_\mathbf{x}(U) \!: U {\supset} E, U \text { open }\} \\ 
    = \delta_\mathbf{x}(X\setminus\{\mathbf{x}\})=0\,,
\end{align}
\end{subequations}
using that singleton sets are closed in Hausdorff spaces, which follows from~\cite[Proposition~4.7]{Folland:1999} noting that it holds for Hausdorff spaces as well, see~\cite[p.~130]{Simmons:topology-analysis}.
This establishes outer regularity of $\delta_{\mathbf{x}}$ on Borel sets. 

Next, we establish inner regularity on open sets. 
Let $\mathbf{x} \in E\in\mathcal{T}$. 
By \cite[4.30 Prop.]{Folland:1999}, there is a compact neighborhood $N$ of $\mathbf{x}$ such that $N \subset E$, since $\X$ is \locallyCompactHausdorff{}, $E \subset \X$ is open, and $\mathbf{x} \in E$.
Thus
\begin{subequations}
\begin{align}
    1=\delta_\mathbf{x}(E)\geq\sup \{\delta_\mathbf{x}(K): K \subset E, K \text { compact }\} \\
    \geq \delta_\mathbf{x}(N)=1\,.
\end{align}
\end{subequations}
Conversely, let $\mathbf{x}\not\in E\in\mathcal{T}$.
Then
\begin{align}
    0=\delta_\mathbf{x}(E)\geq \sup \{\delta_\mathbf{x}(K): K \subset E, K \text { compact }\}\geq 0\,.
\end{align}
This establishes inner regularity on open sets.
As a result, the measure $\delta_\mathbf{x} = \delta_\mathbf{x}^+
+ 0 + 0{\cdot}\mathrm{i}$ is a~complex Radon measure.
\hfill$\square$\\[-0.2cm]

\section{Proof of the Proposed Derivative Being Fréchet (Proposition~\ref{proposition:first-order-Fréchet})}\label{appendix:first-order-Frechet-proof}
The proof is divided into several steps.
First, the first-order directional derivative is treated within a special case in which there are at most two objects present. 
The findings pave the way to the elegant proof of the general case given next. 

\subsection{First-order Derivative and Auxiliary Results}\label{appendix:firtst-order-Frechet:analysis-subsection}
To show that the proposed first-order directional derivative is Fr\'echet, we need to show that the directional derivative $ \frac{\partial G }{\partial \nu} [\eta] $~\eqref{eq:PGFM:directional:derivative} of $G[\eta]$~\eqref{eq:generic-functional-on-M} is a continuous linear operator in the direction $\nu\in M(\X)$, such that
\begin{align}
    \lim_{\|\nu\|\to 0}\frac{\left|G[\eta+\nu]-G[\eta] - \frac{\partial G }{\partial \nu} [\eta] \right|}{\|\nu\|}=0\,,
    \label{app:eq:Frechet:PGFM}
\end{align}
for all $\eta\in M(\X)$, see Supplement~\ref{appendix:derivatives_NLP}.\ref{app:NLP:Frechet}.
The form of the directional derivative could be derived directly from the definition~\eqref{eq:PGFM:directional:derivative}.
It is, however, more convenient to prove that it has the form
\begin{align}
    & \frac{\partial G }{\partial \nu} [\eta] =
    \sum_{n=1}^{+\infty} \frac{1}{n!} \sum_{i=1}^n \overbrace{ \int_\X \dots \int_\X }^{ n\text{-times} } p_{n}(\mathbf{x}_1,\dots,\mathbf{x}_n) \notag\\
    & \hspace{0.6cm} \cdot \eta(\d\mathbf{x}_1) \cdots \eta(\d\mathbf{x}_{i-1})\nu(\d\mathbf{x}_i)\eta(\d\mathbf{x}_{i+1}) \cdots \eta(\d\mathbf{x}_n) \, ,
    \label{app:eq:PGFM:directional-derivative:general}
\end{align}
by showing that it is also Fréchet derivative, i.e., by showing that~\eqref{app:eq:Frechet:PGFM} is true with $\frac{\partial G_{} }{\partial \nu} [\eta]$ having the form~\eqref{app:eq:PGFM:directional-derivative:general}.

The derivative $\frac{\partial G_{} }{\partial \nu} [\eta]$~\eqref{app:eq:PGFM:directional-derivative:general} is defined for any $\nu \in M(\X)$ since all the integrals appearing in~\eqref{app:eq:PGFM:directional-derivative:general} are well defined under the assumption that $p_{n}$ is from $C_0(\X^{n})$.
It is also linear and continuous in~$\nu$ by the \emph{duality} of the linear spaces $M(\X)$ and $C_0(\X)$ discussed in Supplement~\ref{appendix:Radon_Measures}, in particular from the Riesz representation Theorem~\ref{thm:Riesz}~\cite[Theorem~7.17]{Folland:1999}. 

\subsubsection{At Most Two Objects Case}\label{appendix:sub-sec:first-order-Frechet:AtMostTwoObjects}
Consider the case that there are at most two objects in the RFS $\Xi$, i.e., $p_{\Xi}(X)=0$ whenever $|X|\geq 3$.
Indeed,
\begin{subequations}\label{app:eq:G-difference:maxTwoObjects}
\begin{align}
    \!\!\! G_{} [\eta\!+\!\nu]\!-\!G_{} [\eta] 
    &= p_{0} +\textstyle\int_X p_{1} (\mathbf{x}) (\eta\!+\!\nu)(\mathrm{d}\mathbf{x}) \notag\\
    &\hspace{-1.25cm} + \textstyle \frac{1}{2} \int_\X\int_\X p_{2}( \mathbf{x}_1, \mathbf{x}_2)(\eta {+} \nu)(\mathrm{d}\mathbf{x}_1)(\eta {+} \nu)(\mathrm{d}\mathbf{x}_2) \notag\\
    &\hspace{0.35cm} - p_{0} - \textstyle\int_\X p_{1}(\mathbf{x}) \eta(\mathrm{d}\mathbf{x}) \notag\\
    &\hspace{0.35cm} - \textstyle \frac{1}{2} \int_X\int_X p_{2}(\mathbf{x}_1, \mathbf{x}_2) \eta(\mathrm{d}\mathbf{x}_1) \eta(\mathrm{d}\mathbf{x}_2) \\
    & = \textstyle\int_\X p_{1}(\mathbf{x}) \nu(\mathrm{d}\mathbf{x}) \notag\\
    &\hspace{0.35cm} + \textstyle \frac{1}{2} \int_\X\int_\X
    p_{2}(\mathbf{x}_1, \mathbf{x}_2) \eta(\mathrm{d}\mathbf{x}_1) \nu(\mathrm{d}\mathbf{x}_2) \notag\\
    &\hspace{0.35cm} + \textstyle \frac{1}{2} \int_\X\int_\X
    p_{2}(\mathbf{x}_1, \mathbf{x}_2) \nu(\mathrm{d}\mathbf{x}_1) \eta(\mathrm{d}\mathbf{x}_2) \notag\\
    &\hspace{0.35cm} + \textstyle \frac{1}{2} \int_\X\int_\X
    p_{2}(\mathbf{x}_1, \mathbf{x}_2) \nu(\mathrm{d}\mathbf{x}_1) \nu(\mathrm{d}\mathbf{x}_2) . \label{app:eq:G-difference:maxTwoObjects:end}
\end{align}
\end{subequations}
For such a case, Eq.~\eqref{app:eq:PGFM:directional-derivative:general} becomes
\begin{align}
    \tfrac{\partial G_{} }{\partial \nu} [\eta]
    &= \textstyle\int_\X p_{1}(\mathbf{x}) \nu(\mathrm{d}\mathbf{x}) \notag\\
    &\hspace{0.35cm}\textstyle + \frac{1}{2}\int_\X\int_\X
    p_{2}(\mathbf{x}_1, \mathbf{x}_2) \eta(\mathrm{d}\mathbf{x}_1) \nu(\mathrm{d}\mathbf{x}_2) \notag\\
    &\hspace{0.35cm}\textstyle + \frac{1}{2}\int_\X\int_\X
    p_{2}(\mathbf{x}_1, \mathbf{x}_2) \nu(\mathrm{d}\mathbf{x}_1) \eta(\mathrm{d}\mathbf{x}_2) \, ,
\end{align}
and thus~\eqref{app:eq:Frechet:PGFM} becomes
\begin{equation}
    \lim_{\|\nu\| \to 0}
    \frac{ 
        \big|
        \overbrace{
            \textstyle \frac{1}{2}\int_\X\int_\X
            p_{2}(\mathbf{x}_1, \mathbf{x}_2) \nu(\mathrm{d}\mathbf{x}_1) \nu(\mathrm{d}\mathbf{x}_2)
            }^{ G_{} [\eta + \nu] - G_{} [\eta] - \frac{\partial G_{}}{\partial\nu} [\eta] }
        \big|
    }{ \|\nu\| } = 0 \, .
\end{equation}
Recall that $p_{2}(\cdot, \cdot)\in C_0(\X\times\X)$, and that such functions are bounded~\cite[p.~132]{Folland:1999}.
Using $ \| \, |\nu| \, \| = \|\nu\| $, we have
\begin{subequations}\label{app:eq:maxTwoObject:limit:adjustments}
\begin{align}
    &\hspace{-0.0cm} \lim_{\|\nu\|\to 0}
    \frac{\left|\int_\X\int_\X
    p_{2}(\mathbf{x}_1, \mathbf{x}_2) \nu(\mathrm{d}\mathbf{x}_1) \nu(\mathrm{d}\mathbf{x}_2)\right|}{\|\nu\| } \notag\\
    &\leq \lim_{\|\nu\|\to 0} \!\! \frac{ \int_\X \int_\X
    p_{2}(\mathbf{x}_1, \mathbf{x}_2)( |\nu| )(\mathrm{d}\mathbf{x}_1)( |\nu| )(\mathrm{d}\mathbf{x}_2) }{ \| \, |\nu| \, \|^2} \|\nu\| \! \label{app:eq:maxTwoObject:limit:adjustments:three}\\
    &\leq \underbrace{ \sup_{(\mathbf{x}_1, \mathbf{x}_2) \in \X\times\X} p_{2}(\mathbf{x}_1, \mathbf{x}_2) }_{ \hspace{1cm} < +\infty } \cdot \lim_{\|\nu\|\to 0} \|\nu\| \ = \ 0 \, , \label{app:eq:maxTwoObject:limit:adjustments:last}
\end{align}
\end{subequations}
which proves the theorem~\eqref{app:eq:Frechet:PGFM} for the special case of having at most two objects.\\[-0.2cm]

\textit{Remark:}
in general, the densities $p_{n}\in C_0(\X^n)$ are guaranteed to be bounded for any finite $n$~\cite[p.~132]{Folland:1999} since $\X^n$ is \locallyCompactSecondCountableHausdorff{} whenever $\X$ is \locallyCompactSecondCountableHausdorff{} as well (which follows as a special case of the well-known Tychonoff Theorem~\cite[p.~136]{Folland:1999}).
Using the additional uniform bound $K$~\eqref{app:eq:bound-for-all-cardinalities}, any integral of $p_{n}( \mathbf{x}_1,\dots,\mathbf{x}_n ) \in C_0(\X^n) $ taken iteratively with respect to measures $\nu_1,\dots,\nu_n \in M(\X)$ is bounded with
\begin{align}
    \textstyle
    \int_{\X} \cdots\! \int_{\X}
    p_{n}(\mathbf{x}_1,\dots,\mathbf{x}_n) \nu_1( \d\mathbf{x}_1 ) \cdots \nu_n( \d\mathbf{x}_n ) \leq K \prod_{i=1}^n \! \| \nu_i \|  ,
    \label{app:eq:upper-bound-for-iterated-integrals}
\end{align}
using the same procedure as in~\eqref{app:eq:maxTwoObject:limit:adjustments}.
Note that the PGFM $G_{\Xi}[\eta]$~\eqref{eq:def:PGFM} itself is also bounded with 
\begin{align}
    G_{\Xi}[\eta] \leq K \mathrm{e}^{\| \eta \|} \, . 
\end{align}

Using the above arguments, the proof can easily be generalized to the case of having up to any finite number of objects.
Along with this relaxation, the proof is given for any general $m$-th order directional derivative, $m\in\mathbb{N}$.

\subsection{Proof of the $m$-th Order Derivative Being Fréchet}\label{appendix:nth-order-Frechet-proof}
To show that the proposed $m$-th order ($m {\in} \mathbb{N}$) derivative is Fr\'echet, we need to show that the directional derivative $ \frac{\partial^m G_{} }{\partial \nu_1 \cdots \partial \nu_m} [\eta] $~\eqref{eq:PGFM:mth-order-directional:derivative} is a continuous multilinear operator in all directions $\nu_1,\dots,\nu_m\in M(\X)$, such that (cf.~\cite[p.~309]{LusternikSobolev:ElementsOfFunctionalAnalysis:1975})
\begin{align}
    \!\!\!\!\! & \lim_{\|\nu_m\|\to 0} \!\!\!\! \frac{ \left\| \frac{\partial^{m{-}1} G_{} }{\partial \nu_1 \cdots \partial \nu_{m{-}1} }[\eta{+}\nu_m] {-} \frac{\partial^{m{-}1} G_{} }{\partial \nu_1 \cdots \partial \nu_{m{-}1} }[\eta] {-} \frac{\partial^m G_{} }{\partial \nu_1 \cdots \partial \nu_m} [\eta] \right\|_{m\!-\!1} }{\|\nu_m\|} \! \notag\\
    \!\!\!\!\! &\hspace{7cm}=0\,,
    \label{app:eq:Frechet:PGFM:mth}
\end{align}
for all $\eta {\in} M(\X)$.
The norm $\|\cdot\|_{m\!-\!1}$ is a norm on the space of ($m{-}1$)-variate continuous multilinear maps of the form $G_{\nu_m,\eta}(\nu_1,\dots,\nu_{m-1}) \in \mathbb{C}$ that can be written as~\cite[p.~304]{LusternikSobolev:ElementsOfFunctionalAnalysis:1975}
\begin{align}
    \| G_{\nu_m,\eta} \|_{m\!-\!1} &= \!\! \sup_{\nu_1,\dots,\nu_{m-1} \in M(\X) \setminus \{0\} } \!\! \frac{ | G_{\nu_m,\eta}(\nu_1,\dots,\nu_{m-1}) | }{ \| \nu_1 \| \cdots \| \nu_{m-1} \|  } \, ,
    \label{app:eq:norm-on-mulilinear-maps}
\end{align}
where $0$ stands for the measure that assigns zero to any Borel set.
Note that both $\nu_m$ and $\eta$ are fixed in~\eqref{app:eq:norm-on-mulilinear-maps}.

The form of the directional derivative could be derived directly from the definition~\eqref{eq:PGFM:directional:derivative} and its general form is given in Eq.~\eqref{app:eq:PGFM:directional-derivative:general:mth-order}, where $\mathrm{Sym}(m)$ is the set of all permutations on the set $\{1,\dots,m\}$.
\begin{table*}[!ht]
    \rule{\textwidth}{0.4pt}
    General form of the $m$-th order directional derivative~\eqref{eq:PGFM:mth-order-directional:derivative} of the PGFM $G_{\Xi}[\eta]$~\eqref{eq:def:PGFM} in the directions $\eta_1,\dots,\eta_m\in M(\X)$:
    \begin{align}
        \label{app:eq:PGFM:directional-derivative:general:mth-order}
        & \frac{\partial^m G_{} }{\partial \nu_1 \cdots \partial \nu_m} [\eta] =
        \sum_{n=m}^{+\infty} \frac{1}{n!}
        \ \sum_{ 1 \leq i_1 < \dots < i_m \leq n } \
        \sum_{ \pi \in \mathrm{Sym}(m) }
        \overbrace{ \int_\X \cdots \int_\X }^{ n\text{-times} } p_{n}(\mathbf{x}_1,\dots,\mathbf{x}_n) \\ & \hspace{6.0cm}\notag
        \cdot \eta(\d\mathbf{x}_1)
        \cdots \eta(\d\mathbf{x}_{i_1-1}) \nu_{\pi(1)}(\d\mathbf{x}_{i_1}) \eta(\d\mathbf{x}_{i_1+1}) 
        \cdots \eta(\d\mathbf{x}_{i_m-1}) \nu_{\pi(m)}(\d\mathbf{x}_{i_m}) \eta(\d\mathbf{x}_{i_m+1}) 
        \cdots \eta(\d\mathbf{x}_n) .
    \end{align}
\end{table*}
For $m{=}1$, Eq.~\eqref{app:eq:PGFM:directional-derivative:general:mth-order} obviously becomes the first-order derivative~\eqref{app:eq:PGFM:directional-derivative:general} and the following proof thus encompasses~\eqref{app:eq:Frechet:PGFM} as well.

As in Appendix~\ref{appendix:firtst-order-Frechet:analysis-subsection}, it is obvious that the derivative~\eqref{app:eq:PGFM:directional-derivative:general:mth-order} is defined for any $\eta\in M(\X)$, and that it is linear and continuous in all the directions $\nu_1,\dots,\nu_m\in M(\X)$ thanks to the Riesz representation Theorem~\ref{thm:Riesz}~\cite[Theorem~7.17]{Folland:1999}.

\subsection{Proof of Eq.~\eqref{app:eq:Frechet:PGFM:mth}}
First, observe that the term $\frac{\partial^{m{-}1} G_{} }{\partial \nu_1 \cdots \partial \nu_{m{-}1} }[\eta{+}\nu_m]$ given in~\eqref{app:eq:PGFM:term-in-numerator-expanded} can be expanded using a sum that is analogical to the binomial theorem.  
\begin{table*}[!ht]
    \rule{\textwidth}{0.4pt}
    The first term appearing in the numerator of~\eqref{app:eq:Frechet:PGFM:mth} has the form
    \begin{align}
        \label{app:eq:PGFM:term-in-numerator-expanded}
        & \frac{\partial^{m{-}1} G_{} }{\partial \nu_1 \cdots \partial \nu_{m{-}1} }[\eta{+}\nu_m] =
        \sum_{n=m-1}^{+\infty} \frac{1}{n!}
        \ \sum_{ 1 \leq i_1 < \dots < i_{m-1} \leq n } \
        \sum_{ \pi \in \mathrm{Sym}(m-1) }
        \overbrace{ \int_\X \dots \int_\X }^{ n\text{-times} } p_{n}(\mathbf{x}_1,\dots,\mathbf{x}_n) \\ & \hspace{-0.0cm}\notag
        \cdot \underset{ \textcolor{black}{\rule{1.0cm}{0.6pt}} }{\big[ \eta {+} \nu_m \big]} (\d\mathbf{x}_1)
        \cdots \underset{ \textcolor{black}{\rule{1.0cm}{0.6pt}} }{\big[ \eta {+} \nu_m \big]} (\d\mathbf{x}_{i_1-1}) \nu_{\pi(1)}(\d\mathbf{x}_{i_1}) \underset{ \textcolor{black}{\rule{1.0cm}{0.6pt}} }{\big[ \eta {+} \nu_m \big]} (\d\mathbf{x}_{i_1+1}) 
        \cdots \underset{ \textcolor{black}{\rule{1.0cm}{0.6pt}} }{\big[ \eta {+} \nu_m \big]} (\d\mathbf{x}_{i_{m-1}-1}) \nu_{\pi(m)}(\d\mathbf{x}_{i_{m-1}}) \underset{ \textcolor{black}{\rule{1.0cm}{0.6pt}} }{\big[ \eta {+} \nu_m \big]} (\d\mathbf{x}_{i_{m-1}+1}) 
        \cdots \underset{ \textcolor{black}{\rule{1.0cm}{0.6pt}} }{\big[ \eta {+} \nu_m \big]} (\d\mathbf{x}_n).
    \end{align}
    Expanding the \underline{underlined terms} yield a sum of $2^{n-m+1}$ terms, whose $\left( \begin{smallmatrix}
            n{-}m{+}1\\k
        \end{smallmatrix}\right) = \tfrac{(n-m+1)!}{k! (n-m+1-k)!}$ components involve the measure $\nu_m$ precisely $k$-times.
    That is, the sum is analogical to the binomial theorem, which, however, cannot be used directly since complex measures do not commute.
    \rule{\textwidth}{0.4pt}
\end{table*}
As a result, it can be seen that the expression
\begin{align}
    \underbrace{ 
        \tfrac{\partial^{m{-}1} G_{} }{\partial \nu_1 \cdots \partial \nu_{m{-}1} }[\eta{+}\nu_m] {-} \tfrac{\partial^{m{-}1} G_{} }{\partial \nu_1 \cdots \partial \nu_{m{-}1} }[\eta] {-} \tfrac{\partial^m G_{} }{\partial \nu_1 \cdots \partial \nu_m} [\eta]
    }_{ \text{multilinear map abbreviated as } G_{\nu_m,\eta}(\nu_1,\dots,\nu_{m-1}) } \, ,
    \label{app:term-subtraction}
\end{align}
is equal to $\frac{\partial^{m{-}1} G_{} }{\partial \nu_1 \cdots \partial \nu_{m{-}1} }[\eta{+}\nu_m]$~\eqref{app:eq:PGFM:term-in-numerator-expanded} without several terms:
\begin{itemize}
    \item For each $n \geq m {-} 1$, the term $\frac{\partial^{m-1} G_{} }{\partial \nu_1 \cdots \partial \nu_{m-1}} [\eta]$ substracts each summand with $k {=} 0$ from~\eqref{app:eq:PGFM:term-in-numerator-expanded}.
    Since for $n = m {-} 1$, the only possible value of $k$ is $k {=} 0$, the case $n = m {-} 1$ gets canceled entirely. 
    \item For each $n \geq m $, the term $\frac{\partial^m G_{} }{\partial \nu_1 \cdots \partial \nu_m} [\eta]$~\eqref{app:eq:PGFM:directional-derivative:general:mth-order} substracts each summand with $k {=} 1$ from~\eqref{app:eq:PGFM:term-in-numerator-expanded}.
    Since for $n = m $ the only possible value of $k$ left is $k {=} 1$, the case $n = m $ gets canceled entirely as well.
\end{itemize}
By applying an upper bound~\eqref{app:eq:upper-bound-for-iterated-integrals} for each iterated integral in $G_{\nu_m,\eta}(\nu_1,\dots,\nu_{m-1})$ as referenced in~\eqref{app:term-subtraction}, we obtain an analytically tractable expression provided in~\eqref{app:eq:PGFM:term-in-numerator-expanded-bounded}.
\begin{table*}[!ht]
    Upper bound for the absolute value of $G_{\nu_m, \eta}(\nu_1,\dots,\nu_{m-1})$~\eqref{app:term-subtraction}:
    \begin{align}
        \underbrace{ 
            \Big| \tfrac{\partial^{m{-}1} G_{} }{\partial \nu_1 \cdots \partial \nu_{m{-}1} }[\eta{+}\nu_m] {-} \tfrac{\partial^{m{-}1} G_{} }{\partial \nu_1 \cdots \partial \nu_{m{-}1} }[\eta] {-} \tfrac{\partial^m G_{} }{\partial \nu_1 \cdots \partial \nu_m} [\eta] \Big| 
        }_{ | G_{\nu_m,\eta}(\nu_1,\dots,\nu_{m-1}) | }
        \leq \! 
        \sum_{n=m+1}^{+\infty} \!\! \tfrac{1}{n!} \begin{pmatrix} n \\ m{-}1 \end{pmatrix} (m{-}1)! {\cdot} K {\cdot} \Bigg( \sum_{k=2}^{n-m+1} \! \begin{pmatrix} n{-}m{+}1 \\ k \end{pmatrix} \| \eta \|^{(n-m+1)-k} \|\nu_m\|^k \! \Bigg) \! \prod_{i=1}^{m-1} \| \nu_i \| \, ,
        \notag\\[-0.5cm]
        \label{app:eq:PGFM:term-in-numerator-expanded-bounded}
    \end{align}
    \rule{\textwidth}{0.4pt}
\end{table*}
Mere algebraic manipulations within~\eqref{app:eq:PGFM:term-in-numerator-expanded-bounded} suffice to show that 
\begin{align}
    \frac{ | G_{\nu_m,\eta}(\nu_1,\dots,\nu_{m-1}) | }{ \| \nu_1 \| \cdots \| \nu_{m-1} \|  } \leq 
    K \mathrm{e}^{\|\eta\|} \Big( \mathrm{e}^{\|\nu_m\|} - 1 {-} \|\nu_m\| \Big) .
    \label{app:eq:norm-on-mulilinear-maps:adjusted}
\end{align}
Since this bound is independent of $\nu_1,\dots,\nu_{m-1}$, the norm $\| G_{\nu_m,\eta} \|_{m\!-\!1}$~\eqref{app:eq:norm-on-mulilinear-maps} itself is bounded by the right-hand side of~\eqref{app:eq:norm-on-mulilinear-maps:adjusted}.
As a result, we have that
\begin{align}
    \lim_{\|\nu_m\|\rightarrow 0} \!\!\! \tfrac{ \| G_{\nu_m,\eta} \|_{m\!-\!1} }{ \|\nu_m\| } \leq 
    K \mathrm{e}^{\|\eta\|} \!\! \lim_{\|\nu_m\|\rightarrow 0} \! \Big( \tfrac{ \mathrm{e}^{\|\nu_m\|} - 1 }{\|\nu_m\|} {-} 1 \Big)  = 0 , \label{eq:appendix:final-convergence}
\end{align}
by the L'Hôpital rule, which concludes the proof.
\hfill$\square$
\\[-0.2cm]

\textit{Remark:}
It should be noted that the functions involved in the proof $p_n(\mathbf{x}_1,\dots,\mathbf{x}_n) \in C_0(\X^n)$ and thus the densities $p_{\Xi}(\{ \mathbf{x}_1,\dots,\mathbf{x}_n\})$ need neither be generally symmetric nor non\discretionary{-}{-}{-}negative for Proposition~\ref{proposition:first-order-Fréchet} to hold.

Note that complex measures, i.e., from $M(\X)$, are finite and thus $\sigma$-finite.
However, to apply the Fubini--Tonelli Theorem, the considered measures must be positive, i.e., neither complex nor signed.
If the set of the considered measures $\eta,\nu_1,\dots,\nu_m$ does not form a vector space (as it would be the case when restricted to only positive measures), the Fréchet differentiability cannot be established.

Notice that the limit in~\eqref{eq:appendix:final-convergence} is zero \emph{uniformly} with respect to $\eta$ from any bounded subset of $M(\X)$.
Furthermore, the derivative~\eqref{eq:PGFM:mth-order-directional:derivative} can be shown to be continuous with respect to $\eta$ for any $m {\in} \mathbb{N}$ using analogical procedure as in~\eqref{app:term-subtraction}--\eqref{eq:appendix:final-convergence}.
That is, the PGFM $G$ and thus $G_{\Xi}$ is of the class $C^\infty(M(\X))$. 
That is, the Fréchet derivative of $G$ and thus $G_{\Xi}$ at any $\eta \in M(\X)$ of any order $m\in\mathbb{N}$ exists and is continuous as a mapping from $M(\X)$ to the space of $m$\discretionary{-}{-}{-}variate continuous multilinear functionals on $\X$.
This fact may be useful for establishing even stronger results for the chain and product rules, see~\cite[pp.~308--314]{LusternikSobolev:ElementsOfFunctionalAnalysis:1975}.

\ifCLASSOPTIONcaptionsoff
  \newpage
\fi
\bibliographystyle{IEEEtran}
\clearpage
\setcounter{page}{1}

\begin{table*}
    \fontsize{11pt}{20pt}
    \selectfont
    \centering
    {
        \LARGE
        Supplementary Material: Revisiting Functional Derivatives in Multi-object Tracking} \\[0.2cm]
    {
        Jan~Krejčí,~
        Ondřej~Straka,~
        Petr~Girg,~
        and~Jiří~Benedikt
    }\\[0.2cm]
\end{table*}

\setcounter{section}{0}
\renewcommand\appendixname{Supplement}

\section{Relationships of RFSs and Point Processes}\label{appendix:RFS_notes}
\renewcommand{\thesubsection}{\arabic{subsection}}
The relationship among different approaches to modeling point patterns was detailed in a number of publications, see e.g., ~\cite[Proposition~5.3.II]{Vere-Jones:2003},\cite{Vo-Singh-Bayes:2004,Vo-Singh-Doucet_Sequential-RFS:2005,Mahler-Book:2007,Mahler-Book:2014,MoriChongChang:ThreeFormalism:2016,Baudin:1984}.
This appendix summarizes the relationships among various density functions used to describe RFSs.

An unordered set of $n$~points can be described using a joint distribution.
In order to describe an (unordered) set of points, the joint distribution $\Pi_n(S_1 \times \dots \times S_n)$ should be symmetric~\cite[p.~124]{Vere-Jones:2003}.
If not, it can be symmetrized to yield
\begin{align}\label{eq:symmetrized_distribution}
    \Pi_n^{\mathrm{sym}}(S_1 \! \times \! \dots \! \times \! S_n) = \tfrac{1}{n!} \textstyle \sum_{\sigma} \Pi_n(S_{\sigma(1)} \! \times \! \dots \! \times \! S_{\sigma(n)}),
\end{align}
where the summation goes over all permutations $\sigma$ of the set of $n$ elements.
The measure $\Pi_n^{\mathrm{sym}}$~\eqref{eq:symmetrized_distribution} defines the Janossy measure as~\cite[pp.~124]{Vere-Jones:2003}
\begin{align}
    J_n(S_1 \times \! \dots \! \times S_n) = n! \cdot p_n \Pi_n^{\mathrm{sym}}(S_1 \times \!\dots\! \times S_n),
\end{align}
where $p_n$ is the probability that there are $n$ points in the set.

As noted in Section~\ref{sec:preliminaries}, in point process theory, a \emph{probability} density can be defined as a Radon--Nikodým derivative.
In the notation of this paper, it could be defined as the derivative of $P_\Xi(C)$~\eqref{eq:probability_measure_RFS}, for example w.r.t.~the unnormalized measure of the totally finite unit-rate Poisson point process~\cite{Vo-Singh-Bayes:2004}
\begin{align}\label{eq:unnormalized_unit-rate-PPP_measure}
    P_{\mathrm{Poi}}^{\mathrm{u}}(C) = \sum_{n=0}^{+\infty} \frac{ \referenceMeasure^{n}(\chi^{-1}(C) \cap \X^n ) }{ u^{n} \cdot n! } ,
\end{align}
whose unit of measurement is taken to be $u^{-1}$ and where $\chi$ is the mapping from vectors to sets defined as $\chi( [\mathbf{x}^1,\dots,\mathbf{x}^n]\T ) = \{\mathbf{x}^1,\dots,\mathbf{x}^n\}$.
If the space $(\X,\B,\referenceMeasure)$ is such that $\referenceMeasure(\X)<+\infty$, e.g., $\X$ is a bounded subset of $\mathbb{R}^2$, one could also use the normalized version of \eqref{eq:unnormalized_unit-rate-PPP_measure} as $P_{\mathrm{Poi}}(C) = \mathrm{e}^{-\referenceMeasure(\X)} P_{\mathrm{Poi}}^{\mathrm{u}}(C)$ \cite[Sec.~10.4]{Vere-Jones:2008}.
Relations among various density functions, if they exist, are \cite[Sec.~5.3]{Vere-Jones:2003},\cite{Vo-Singh-Bayes:2004}
\begin{subequations}
\begin{align}
    \frac{\d P_\Xi}{\d P_{\mathrm{Poi}}^{\mathrm{u}} } (X)
    &= \mathrm{e}^{-\referenceMeasure(\X)} \frac{\d P_\Xi}{\d P_{\mathrm{Poi}} } (X) \\
    &= u^{|X|} \cdot |X|! \cdot p_{|X|} \cdot \frac{\d \Pi_{|X|}^{\mathrm{sym}} }{\d \referenceMeasure^{|X|} } (\chi^{-1}(X)) \\
    &= u^{|X|} \cdot \hspace{-0.8cm} \underbrace{ \frac{\delta \beta_\Xi}{\delta X}(X) }_{p_\Xi(X) \ = \ j_{|X|}( \chi^{-1}(X) ) },
\end{align}
\end{subequations}
where $X\in\mathcal{F}(\X)$ is a finite set.
It can be seen that the Janossy densities $j_n(\mathbf{x}_1,\dots,\mathbf{x}_n)$ and thus $p_\Xi(X)$~\eqref{eq:PDF_from_BMF} are not \emph{probability} densitites, but they are related to the probability density $\frac{\d P_\Xi}{\d P_{\mathrm{Poi}}^{\mathrm{u}} }(X)$ via a simple equation. 
Observe that the main concern in this inconvenience is the unit of measurement $u$.

\section{Set Derivatives}\label{appendix:set_derivatives}
The set derivative \emph{constructs} the Radon--Nikodým derivative point-wise in $\X$.
Eventually, it is a Lebesgue density theorem \emph{specified} by using a sequence of closed balls of radius $\tfrac{1}{i}$ in the limit for $i\rightarrow +\infty$.
When applied to a BMF of an RFS, it turns out to form a counterpart of the set integral, i.e., to form a fundamental theorem of multi-object calculus.

Let $\beta(S)$ be real, or possibly multivariate (vector-valued) unitless function defined on closed subsets $S\subseteq \X$.
To be consistent with the setting of this paper, the original definition of the set derivative~\cite[pp.~144--150]{Mahler:1997} can be written as\footnote{
    In~\cite{Mahler:1997}, it was assumed that the base space is $\X=\mathbb{R}^d \times W$, where $W$ is a finite set.
    Although the general case of assuming $\X$ to be \emph{any} \locallyCompactSecondCountableHausdorff{} space may appear in the literature~\cite{Mahler-Book:2014}, the full (i.e., not simplified) definition of set derivatives for this general case is, to the best of our knowledge, unavailable.
}
\begin{align} \label{eq:set_derivatives_Original}
    \frac{\delta \beta}{\delta \mathbf{x}}(S) \!=\! \lim_{j\rightarrow\infty} \lim_{i\rightarrow\infty} \frac{ \beta\big( (S \!\setminus\! F_{\mathbf{x},j}) \cup E_{\mathbf{x},i} \big) - \beta( S \!\setminus\! F_{\mathbf{x},j} ) }{ \referenceMeasure( E_{\mathbf{x},i} ) }  ,
\end{align}
where $\{ E_{\mathbf{x},i} \}_{i=1}^{+\infty}$ is a sequence of \emph{closed} balls converging to~$\{\mathbf{x}\}$, and $\{ F_{\mathbf{x},j} \}_{j=1}^{+\infty}$ is a sequence of \emph{open} balls whose closures converge to $\{\mathbf{x}\}$.
Note that $\referenceMeasure$ is expressed in the units of measurement denoted with~$u$, and thus if $\beta$ is unitless (a probability), the unit of measurements of~\eqref{eq:set_derivatives_Original} is $u^{-1}$.

A simplified definition can be given when $\mathbf{x}\cap S = \emptyset$, in which case \eqref{eq:set_derivatives_Original} could be heuristically expressed as~\cite{Mahler-PHD:2003}
\begin{align}\label{eq:def_Set_derivatives_simplified}
    \frac{\delta \beta}{\delta \mathbf{x}}(S) = \lim_{ \referenceMeasure(E_{\mathbf{x}})\rightarrow 0} \frac{ \beta(S\cup E_{\mathbf{x}}) - \beta(S) }{ \referenceMeasure(E_{\mathbf{x}}) } \, ,
\end{align}
where $E_{\mathbf{x}}$ is simply a neighbourhood of $\mathbf{x}$.
Iterated derivatives w.r.t.~a finite set $X\subset \X$ are given in the same manner as that of iterated functional derivatives~\eqref{eq:def_functional_derivative_iterated}, i.e.,
\begin{subequations}
\begin{align}
    \frac{\delta \beta }{\delta \emptyset}(S) &= \beta(S) \, , \\
    \frac{\delta \beta }{\delta \{ \mathbf{x}_1, \dots, \mathbf{x}_n \} }(S) &= \frac{\delta}{\delta \mathbf{x}_n } \frac{\delta \beta }{\delta \{ \mathbf{x}_1, \dots, \mathbf{x}_{n-1} \} }(S) \, .
\end{align}
\end{subequations}

Note that the set derivative operates on set functions, and sets \emph{do not} form a vector space, i.e., sets cannot be added or scaled.
The following Appendix deals with the case when the domain of the function to be differentiated \emph{does} form a vector space.

\section{Derivatives of a Functional}\label{appendix:derivatives_NLP}
This section summarizes the notion of derivatives in abstract spaces, namely the space of functions or measures.
Let $\mathcal{Y}$ be a vector (i.e., linear) space and $G:\mathcal{Y}\rightarrow\mathbb{C}$, i.e., a functional.
The \emph{directional derivative}\footnote{
    Note that Eq.~\eqref{eq:def_directional_derivative} can be defined more generally, and in the case $G$ is a functional, it is usually called the \emph{variational} derivative~\cite[Appendix~C.3]{Streit-Book:2021}.
} of $G[h]$ in the direction of $g\in\mathcal{Y}$ is given by~\cite[p.~293]{LusternikSobolev:ElementsOfFunctionalAnalysis:1975}
\begin{align}
    \frac{\partial G[h]}{\partial g} \! \triangleq \!
    \lim_{\epsilon \rightarrow 0} \frac{G[h \! + \! \epsilon \! \cdot \! g] - G[h] }{\epsilon} 
    \! = \! \left[ \frac{\d}{\d \epsilon} G[h \! + \! \epsilon \! \cdot \! g] \right]_{\epsilon = 0}
    \label{eq:def_directional_derivative}
     , 
\end{align}
if it exists for the particular function $g$.
Note that it is a common misconception to \emph{call}~\eqref{eq:def_directional_derivative} a Gâteaux or Fréchet derivative: such derivatives require further properties that are not granted automatically.
If the derivative~\eqref{eq:def_directional_derivative} exists for every $g\in\mathcal{Y}$, it is called the \emph{Gâteaux differential}.
If it is moreover linear and continuous as a mapping
\begin{align}
    g \mapsto \frac{\partial G[h]}{\partial g} \, ,
\end{align}
it is called the \emph{Gâteaux derivative}.
Note that linearity of the derivative w.r.t.~the argument $h$ is not required~\cite[p.~295]{LusternikSobolev:ElementsOfFunctionalAnalysis:1975} and it would be unapplicable for functionals that are more complicated than those of a ``quadratic type''.

To be useful, a differential/derivative is required to yield differentiation rules.
As discussed before, for MOT problems, the chain rule is essential~\cite{clark2012faa,Mahler_GeneralizedPHD:2012}.
It can be granted by posing further requirements on the derivative as follows.

\subsection{Normed Linear Spaces: Fréchet derivative}\label{app:NLP:Frechet}
Assume the underlying space is a normed space $(\mathcal{Y},\|\cdot\|)$, e.g., a space of functions or measures.
If~\eqref{eq:def_directional_derivative} is a Gâteaux derivative, and it moreover satisfies~\cite[p.~292]{LusternikSobolev:ElementsOfFunctionalAnalysis:1975}
\begin{align}
    \lim_{\|g\|\rightarrow 0} \frac{ \big| G[h+g] - G[h] - \frac{\partial G[h]}{\partial g} \big| }{\| g \|} = 0 \, , \label{eq:frechet_derivative}
\end{align}
the derivative is further called the \emph{Fréchet derivative} and it yields the chain and product rules~\cite[pp.~312--313]{LusternikSobolev:ElementsOfFunctionalAnalysis:1975}.
Note that Fréchet derivatives are standard in mathematics, but for the chain rule to hold \cite{Bernhard:ChainDifferential:2005}, they might pose too restrictive assumptions on the functional.

\subsection{Topological Vector Spaces: Chain Differential}\label{appendix:Chain-differential}
Assume the underlying space is a topological vector space, e.g., a set of measures equipped with a topology but not with a norm, or the set of distributions which does \emph{not} form a normed space whatsoever.
If the Gâteaux differential $\frac{\partial G[h]}{\partial g}$ satisfies
\begin{align}
   \lim_{i\rightarrow +\infty} \frac{G[h+\theta_i h_i] - G[h]}{\theta_i} = \frac{\partial G[h]}{\partial g} \, ,
\end{align}
for any sequence $\{h_i\}_{i=1}^{+\infty}\subset \mathcal{Y}$ converging to $g$ and any sequence $\{\theta_i\}_{i=1}^{+\infty}$ of nonzero real numbers converging to zero, the differential is called the \emph{Chain differential}~\cite{Bernhard:ChainDifferential:2005}.
If a Gâteaux differential is moreover jointly continuous in both $h$ and the direction $g$, it is also a continuous Chain differential \cite[Corollary~1]{Bernhard:ChainDifferential:2005}, and it follows therein that the chain rule can be readily used.
\\[-0.2cm]

Generally, there is no notion of \emph{gradient} in abstract spaces.
Historically, the first approach to define gradient for functionals, i.e., functional derivative, could be coined to the \emph{Volterra derivative}~\cite[Sec.~2.1]{Volterra:funcionalCalcullus:1930}.
Note that the calculus of variations~\cite{Liberzon:Variations_OptimalControl:2012}, \cite[Appendix~C.3]{Streit-Book:2021} does not define the functional derivative explicitly as it is not needed, e.g., for solving optimization problems.
Assume, on the other hand, there exists a functional transformation $T[h](\mathbf{x})$, using which the directional derivative can be written $\forall g\in\mathcal{Y}$ as\footnote{\label{footnote:functional-on-Rd}
    In the case $\mathcal{Y} = \mathbb{R}^{d}$, the functional $G[h]$ becomes a multivariate function indexed by $\mathbf{x}$, $g$ becomes a direction vector, and the integral restorts to a sum, thus~\eqref{eq:implicit_definition_variations} shrinks to the familiar relation known from undergraduate calculus.
}
\begin{align}
    \frac{\partial G[h]}{\partial g} = \int T[h] (\mathbf{x}) g(\mathbf{x}) \d \mathbf{x} \, . \label{eq:implicit_definition_variations}
\end{align}
Then the existence of the functional derivative $T[h](\mathbf{x})$ can be established using, e.g., the Riesz representation theorem (which does not define $T[h](\mathbf{x})$ point-wise).
However, it is not guaranteed that the directional derivative of a general $G[h]$ can be expressed in the form of~\eqref{eq:implicit_definition_variations}, see~\cite[pp.~11]{Gelfand-Fomin:Variations:1963}, \cite{DensityFunctionalTheory:2011}.

\section{Locally Compact, Second Countable Hausdorff Topological Space}\label{appendix:topology}
Topological spaces are primarily motivated by the need for describing \emph{continuity} of functions defined over elements $\mathbf{x} {\in} \X$ of various nature.
A~topological space $(\X, \mathcal{T})$ is a~pair, where $\X$ is a nonempty set and $\mathcal{T}$ is a set of all \emph{open subsets} of $\X$, further called \emph{topology}\footnote{
    Any topology $\mathcal{T}$ satisfies $\emptyset, \X\in \mathcal{T}$ and $\mathcal{T}$ is closed under arbitrary unions and finite intersections of its elements.
    Unless $\X$ is a singleton, it can be endowed with different topologies.
}~\cite[Chap.~4]{Folland:1999}.
If the topology $\mathcal{T}$ is fixed during the discourse and no confusion may arise, the topological space is denoted only with $\X$ for convenience. 

In the following, the properties ``\emph{locally compact}'', ``\emph{second countable}'' and ``\emph{Hausdorff}'' are described, and convenient sets of functions on $\X$ that are \emph{continuous} w.r.t.~$\mathcal{T}$ are introduced.

\subsection{Definition and Role of \locallyCompactHausdorff{} and \locallyCompactSecondCountableHausdorff{} Spaces}
In general, a set $M\subset \X$ is called \emph{closed} in $(\X, \mathcal{T})$, if $\X {\setminus} M$ is open, i.e., if $(\X {\setminus} M) \in \mathcal{T} $.
Now, let $A \subset \X$ be an arbitrary set.
By \emph{interior} of $A$, denoted by $A^{\circ}$, we mean the union of all open sets contained in $A$.
By \emph{closure} of $A$, denoted by $\bar{A}$, we mean the intersection of all closed sets containing $A$. 
A set $U\subset \X$ is called a \emph{neighbourhood} of $x\in\X$, if $x\in U^{\circ}$.

\textbf{Hausdorff property:}
distinct points of a \emph{Hausdorff} space (also called T$_2$) have disjoint neighbourhoods.
Formally, a topological space is called \emph{Hausdorff},
if, for any two points $\mathbf{x},\mathbf{y}\in\X$, with $\mathbf{x} \not= \mathbf{y}$, there exist open sets
$U, V$ with $\mathbf{x} {\in} U$, $\mathbf{y} {\in} V$ such that
$U\cap V=\emptyset$, cf. \cite[pp.~116--117]{Folland:1999}.

\textbf{Local compactness:}
first, \emph{compactness} generalizes the notion of closed-and-bounded sets.
Formally, a set $Y \subset X$ is called {\it compact}, if for any $\left\{U_\alpha\right\}_{\alpha \in A}\subset \mathcal{T}$ such that $Y \subset \bigcup_{\alpha \in A} U_\alpha$, there exists a \emph{finite} subset $B \subset A$ such that $Y \subset \bigcup_{\alpha \in B} U_\alpha$,
\cite[p.~128]{Folland:1999}.
A topological space is called {\it locally compact} if every point has a compact neighborhood.

\textbf{Second countability:}
first, a \emph{base} $\mathcal{B} {\subset} \mathcal{T}$ for the topology~$\mathcal{T}$ is a collection of open sets that can ``recover'' the topology $\mathcal{T}$ and thus any nonempty open set $U$, so that~\cite[p.~115]{Folland:1999}
\begin{align}
    \mathcal{T} &= \Big\{ \underbrace{ \cup_{B \in \beta} B }_{U} \Big| \beta \subset \mathcal{B} \Big\} \, .
\end{align}
A topological space is called \emph{second countable} (aka \emph{completely separable}) if the topology has \emph{countable base}, see~\cite[p.~116]{Folland:1999}.

\emph{Locally compact Hausdorff} (\locallyCompactHausdorff{}) spaces play important role in analysis~\cite[Sec.~4.5]{Folland:1999} and the additional assumption of \emph{second countability} is useful when dealing with products of Radon measures, see, e.g.~\cite[Sec.~7.4]{Folland:1999}. 
Further, as discussed before, \locallyCompactSecondCountableHausdorff{} spaces are used in the theory of random (finite) sets.
If a \locallyCompactSecondCountableHausdorff{} space is moreover a vector space, it is of finite dimension due to the local compactness assumption~\cite[Theorem~1.22]{Rudin:FunctionalAnalysis:1973}.
Also note that a \locallyCompactSecondCountableHausdorff{} space is a \emph{Polish} space, i.e., it can be turned into a \emph{complete separable metric} (\completelySeparableMetric{}) space, see~\cite[Theorem~5.3]{Kechris:DescriptiveSetTheory:1995}.
The assumption of the space being \completelySeparableMetric{} is usual in the point process literature \cite{Vere-Jones:2008}. 

\subsection{Convenient Sets of Functions on Topological Spaces}
Let $(\X_1, \mathcal{T}_1)$ and $(\X_2, \mathcal{T}_2)$ be two topological spaces.
A~function $f\colon \X_1\to \X_2$ is called \emph{continuous}, if the pre-image\footnote{
    By pre-image we mean $f^{-1}(B):=\{x\in \X_1\colon f(x)\in B\}$.
} $f^{-1}(B)$ is from $\mathcal{T}_1$ for any set $B\in\mathcal{T}_2$~\cite[p.~119]{Folland:1999}.
When $\X_2$ is $\mathbb{R}$ or $\mathbb{C}$, the topology $\mathcal{T}_2$ is assumed to be the \emph{standard} topology induced by the usual metric on $\mathbb{R}$ or $\mathbb{C}$, i.e., by $d(x,y)=|y-x|$.

Let $(\X, \mathcal{T})$ be a topological space and $C(\X)$ denote the set of all complex-valued functions on $\X$ that are continuous w.r.t.~the topology $\mathcal{T}$.
For $f \in C(\X)$, let the \emph{closed} set
\begin{align}
    \operatorname{supp}(f) \triangleq \overline{  \{ \mathbf{x} \in \X\colon f(\mathbf{x}) \neq 0\} } \, ,
    \label{app:support-topology}
\end{align}
be called a \emph{support} of $f$ \emph{in the topology} $\mathcal{T}$.
Next, denote the space of \emph{compactly supported} functions of $C(\X)$ as
\begin{align}
    C_c(\X) \triangleq \{f \in C(X): \operatorname{supp}(f) \text { is compact }\} \, .
\end{align}
We say that $f \in C(\X)$ \emph{vanishes at infinity}, if for every $\epsilon>0$ the set $\{ \mathbf{x} \in \X:|f( \mathbf{x} )| \geq \epsilon\}$ is compact.
The corresponding function space is denoted by
\begin{align}
    C_0(\X) \triangleq \{f \in C(\X): f \text { vanishes at infinity }\} \, .
\end{align}
Clearly, $C_c(\X)$ is a subspace of $C_0(\X)$, which is a subspace of $C(\X)$.
Note that $C_0(\X)$ forms a normed vector space using the \emph{uniform norm} defined as~\cite[p.~121]{Folland:1999}
\begin{align}
    \| f \| = \sup_{x\in \X} \, |f(x)| \, ,
\end{align}
which is required in Theorem~\ref{thm:Riesz}.
The above sets of continuous functions may be too small for general topological spaces~\cite[p.~121]{Folland:1999}, but for \locallyCompactSecondCountableHausdorff{} spaces they are \emph{rich enough} and have \emph{nice properties} further discussed in the following appendix.

\section{Radon Measures and Dual of $C_0(\X)$}\label{appendix:Radon_Measures}
At this point, we assume that the reader is familiar with definitions of a {\it measure}~\cite[Sec.~1.3, p.~24]{Folland:1999}, a {\it signed measure}~\cite[Sec.~3.1, p.~85]{Folland:1999}, and a {\it complex measure}~\cite[Sec.~3.3, p.~93]{Folland:1999} on a {\it measure space} $(\X, \Sigma)$, where $\Sigma$ stands for a $\sigma$-{\it algebra}~\cite[Sec.~1.2, p.~21]{Folland:1999} of subsets of $\X$. 

Let $(\X, \mathcal{T})$ be a topological space. 
Then $\sigma$-algebra generated by $\mathcal{T}$ is called
\emph{Borel $\sigma$-algebra} and it is denoted by $\mathfrak{B}(\X, \mathcal{T})$. Elements of $\mathfrak{B}(\X, \mathcal{T})$ are called \emph{Borel subsets} of~$\X$. A measure defined on all Borel subsets of $\X$ is called \emph{Borel measure}.
For \locallyCompactSecondCountableHausdorff{} spaces, Borel measures are well-behaved with the following properties.

Let $\eta$ be a Borel measure on $\X$ and $E$ a Borel subset of~$\X$.
The measure $\eta$~is called \emph{outer regular} on $E$ if
\begin{align}
    \eta(E)=\inf\{\eta(U): U \supset E,\ U \text{ open }\} \, ,
\end{align}
and \emph{inner regular} on $E$ if
\begin{align}
    \eta(E)=\sup \{\eta(K): K \subset E,\ K \text{ compact }\} \, .
\end{align}
Following~\cite[p.~212]{Folland:1999}, the Borel measure $\eta$ is called \emph{Radon measure}, if it is \emph{(i)} finite on all compact sets, \emph{(ii)} outer regular on all Borel sets, and \emph{(iii)} inner regular on all open sets.

According to~\cite[p.~222]{Folland:1999}, a \emph{signed} Radon measure is a \emph{signed} Borel measure whose positive and negative \emph{variations} (see below) are Radon.
Further, a \emph{complex} Radon measure is a \emph{complex} Borel measure, i.e., taking values in $\mathbb{C}$, whose real and imaginary parts are signed Radon measures. 
Unlike for (real) measures, note that infinite values are not allowed for complex measures.

The set $M(\X)$ of all complex Radon measures on $\X$ forms a normed linear space over the field $\mathbb{C}$, where the norm for each $\eta \in M(\X)$ is defined as
\begin{align}
    \|\eta\|=|\eta|(\X) \, ,
\end{align}
where $|\eta|(E)$ for any $E\in\mathfrak{B}(\X,\mathcal{T})$ is the \emph{total variation} of~$\eta$.
For $\eta$ being a (real) signed measure, the total variation of $\eta$ is a measure defined as
\begin{align}
    |\eta|(E) \triangleq \eta^+(E) + \eta^-(E) \, ,
\end{align}
with $\eta^+$ and $\eta^-$ being the positive and negative variations (both are positive measures) given by the Jordan decomposition theorem~\cite[pp.~87--88]{Folland:1999}.
For $\eta$ being a complex measure, the total variation is the sum of the total variations of its real and imaginary parts~\cite[p.~93]{Folland:1999}.

The following theorem is an essential tool in establishing our results in Section~\ref{sec:proposed}, namely to show that the proposed directional derivative is continuous in the direction $\nu \in M(\X)$. 

\begin{theorem}
\label{thm:Riesz}
(The Riesz Representation Theorem~\cite[Theorem~7.17]{Folland:1999})
Let $\X$ be \locallyCompactHausdorff{}, and for $\eta {\in} M(\X)$ and $f {\in} C_0(\X)$ let $I_\eta(f) = \! \int \! f d \eta$. Then the map $\eta \mapsto I_\eta$ is an isometric isomorphism from $M(\X)$ to $C_0(\X)^*$, where $C_0(\X)^*$ is the (topological) dual space of $C_0(\X)$~\cite[p.~157]{Folland:1999}. 
\end{theorem}

That is, the space $M(\X)$ of all complex Radon measures on $\X$ can be identified with the space $C_0(\X)^*$ of all bounded linear functionals on $C_0(\X)$.

\section{Dirac Deltas}\label{appendix:Dirac_deltas}
There is significant demand in engineering and physics for methods to handle signals and densities of quantities that converge to a single point in the limit, while maintaining a constant total yield of the given quantity throughout this limiting process.
In this paper, the shrunk quantities are \emph{Dirac deltas}.
They are used to mimic \emph{standard basis vectors} to be used as \emph{directions} for computing derivatives of functionals.

As a typical example, see e.g.,~\cite{Streit_FunctionalDerivatives:2015}, consider function sequences such as $f_n(x){=}\tfrac{n}{2}$ for $x\in[-1/n, 1/n]$
and $f_n(x){=}0$ elsewhere in $\mathbb{R}$ for $n\to\infty$.
Note that $\int_{\mathbb{R}} f_n(x)\mathrm{d}x {=} 1$ for all $n {\in} \mathbb{N}$.
One may ask \emph{in what sense} one needs to consider $\lim_{n\to\infty}f_n$, because the pointwise limit of this function sequence is the \emph{function} $f$ with $f(0) {=} {+} \infty$ and $f(x) {=} 0$ elsewhere in $\mathbb{R}$.
Although there is a constant \emph{yield} throughout the limiting process $\lim_{n\to\infty}\int_{\mathbb{R}} f_n(x)\mathrm{d}x {=} 1$, we see that the pointwise limit function $f$ does \emph{not} retain the yield since $\int_{\mathbb{R}}f(x)\mathrm{d}x {=} 0 \not= 1$.
To capture the constant yield within the limiting entity, it turns out that the entity cannot be a \emph{standard} function, see the example given in~\cite[pp. 376--377, Remark 14]{Mahler-Book:2007}.
In the following, two possibilities of how the limiting process can be considered are discussed.

\subsection{Dirac Distribution}\label{appendix:Dirac_distribution}
Distributions are continuous linear functionals defined on the topological vector space $C_c^{\infty}(\X)$~\cite{Folland:1999} with $\X \subseteq \mathbb{R}^d$ being an \emph{open subset}, called the space of \emph{test functions}~\cite{Folland:1999}.
Notice that the space $\X$ must be such that one can conveniently tackle derivatives of functions defined on it, i.e, not an arbitrary \locallyCompactHausdorff{} or \locallyCompactSecondCountableHausdorff{} space, see~\cite[Chapter~9]{Munkers:AnalysisOnManifolds:1991}.
The \emph{Dirac distribution} $\delta_{\mathbf{x}}:C_c^{\infty}(\X) \rightarrow \mathbb{C}$ centered at $\mathbf{x}\in\X$, can be defined as
\begin{align}
    \delta_{\mathbf{x}}[\varphi]=\varphi(\mathbf{x}) \, , \label{dirac-delta-distribution}
\end{align}
for all $\varphi\in C_c^{\infty}(\X)$.
The linear transformation $\delta_{\mathbf{x}}$~\eqref{dirac-delta-distribution} can be \emph{approximated} via members of a 
family $\{ \DiracTestSeq{\lambda}{\mathbf{x}}(\cdot) \}_{\lambda}$ of certain functions indexed by $\lambda {>} 0$, such that the following \emph{approximate identity} can be used to characterize $\delta_{\mathbf{x}}$~\eqref{dirac-delta-distribution} as
\begin{align}
    \lim_{\lambda \searrow 0} \int_{\X} \DiracTestSeq{\lambda}{\mathbf{x}}(\mathbf{y}) \varphi(\mathbf{y}) \d \mathbf{y} = \varphi(\mathbf{x}) \, , \label{eq:family-of-functions:distributions:appendix}
\end{align}
which must be valid for all \emph{test functions} $\varphi\in C_c^{\infty}(\X)$, and the convergence of $\DiracTestSeq{\lambda}{\mathbf{x}}(\cdot)$ to $\delta_{\mathbf{x}}$ then concerns the topology in $C_c^{\infty}(\X)$.
This is a simple consequence of~\cite[Proposition~9.1]{Folland:1999} using the translation property of the Dirac delta~\cite[p.~285]{Folland:1999}.
The family $\{\DiracTestSeq{\lambda}{\mathbf{x}}\}_{\lambda}$ can be specified by setting \mbox{$\DiracTestSeq{\lambda}{\mathbf{x}}(\mathbf{y})=\lambda^{-d}f(\tfrac{ \mathbf{y}-\mathbf{x} }{ \lambda })$},
using \emph{any} absolutely integrable function $f$ on $\mathbb{R}^d$ for which $\int_{\mathbb{R}^d} f \d \referenceMeasure {=} 1$.
For instance, $f$ can be chosen as a Gaussian or indicator function~\cite{Streit_FunctionalDerivatives:2015}.
Note that~\eqref{eq:family-of-functions:distributions:appendix} alone is valid for functions $\varphi$ from a larger set of functions than $C_c^{\infty}(\X)$, see~\cite[Theorem~8.14b]{Folland:1999}, \cite[Theorem~8.15]{Folland:1999}, which, however, do not posses the same topology. 
That is, the calculus of distributions requires $\varphi \in C_c^{\infty}(\X)$, for the Dirac distribution $\delta_{\mathbf{x}}$~\eqref{dirac-delta-distribution} to be the limit in~\eqref{eq:family-of-functions:distributions:appendix}, as it would lead to problems when considered with certain functions $\varphi \in B(\X) \setminus C_c^{\infty}(\X)$. 

\subsection{Dirac Measure}\label{appendix:Dirac_measure}
Let $(\X, \Sigma)$ be any measurable space.
The \emph{Dirac measure} $\delta_{\mathbf{x}}: \Sigma \rightarrow \{0,1\}$ centered at $\mathbf{x}\in\X$ is a measure defined by 
\begin{align}
    \delta_\mathbf{x}(E):= \begin{cases}0 \, , & \mathbf{x} \notin E \, , \\ 1 \, , & \mathbf{x} \in E \, , \end{cases} \label{eq:Dirac-measure-def}
\end{align}
for any set $E\in\Sigma$.
For~\eqref{eq:Dirac-measure-def} to be well defined, note that $\Sigma$ could be any $\sigma$-algebra on $\X$, even the power set.
 
Note that the map $I_{\delta_\mathbf{x}}\colon C_0(\X)\to \mathbb{C}$ defined as
\begin{align}
    I_{\delta_\mathbf{x}}(f)=\int_X f \mathrm{d}\delta_\mathbf{x} = f(\mathbf{x}) \, ,
\end{align}
for any $f\in C_0(\X)$ and any given $\mathbf{x}\in\X$ is \emph{continuous} by the Riesz representation Theorem~\ref{thm:Riesz}~\cite[Theorem~7.17]{Folland:1999}, since $\delta_\mathbf{x}$ is a~complex Radon measure.

\end{document}